\documentclass[aps,prd,twocolumn,groupedaddress,showkeys,floatfix,nofootinbib]{revtex4}

\usepackage[T1]{fontenc}

\usepackage{graphicx}	
\usepackage{subcaption}
\usepackage{booktabs}
\usepackage{newtxtext,newtxmath}
\usepackage{multirow}
\usepackage{array}
\usepackage{aas_macros}

\usepackage{xcolor}
\definecolor{xlinkcolor}{cmyk}{1.0,0.7,0,0}

 \usepackage[bookmarks=true,         
     pdfnewwindow=true,      
     colorlinks=true,    
     linkcolor=xlinkcolor,     
     citecolor=xlinkcolor,     
     filecolor=xlinkcolor,  
     urlcolor=xlinkcolor,      
     final=true,
 ]{hyperref}

\newcommand{\pder}[2]{\frac{\partial#1}{\partial#2}}

\def\BV/{Brunt-V\"{a}is\"{a}l\"{a}}

\begin{document}


\title{Remnant masses from 1D+ core-collapse supernovae simulations: bimodal neutron star mass distribution and black holes in the low-mass gap}


\author{Luca Boccioli}
\email{lbocciol@berkeley.edu}
\affiliation{Department of Physics, University of California, Berkeley, CA 94720, USA}

\author{Giacomo Fragione}
\affiliation{Center for Interdisciplinary Exploration and Research in Astrophysics (CIERA), 1800 Sherman, Evanston, IL 60201, USA}
\affiliation{Department of Physics \& Astronomy, Northwestern University, Evanston, IL 60208, USA}


\date{\today}

\begin{abstract}
The explosion of core-collapse supernovae (CCSNe) is an extremely challenging problem, and there are still large uncertainties regarding which stars lead to successful explosions that leave behind a neutron star, and which ones will form a black hole instead. In this paper, we simulate 341 progenitors at three different metallicities using spherically symmetric simulations that include neutrino-driven convection via a mixing-length theory. We use these simulations to improve previously derived explosion criteria based on the density and entropy profiles of the pre-supernova progenitor. We also provide numerical fits to calculate the final mass of neutron stars based on either compactness, the location of the Si/Si-O interface, or the Chandrasekhar mass. The neutron star birth mass distribution derived from our 1D+ simulations is bimodal, contrary to what the most popular 1D CCSN simulations have shown so far. We compare the theoretically derived neutron star mass distributions with the observed ones and discuss potential implications for population synthesis studies. We also analyze the black hole mass distribution predicted by our simulations. To be consistent with current models of matter ejection in failed SNe, a large fraction of the envelope must be expelled, leading to small black holes in the low-mass gap. One black hole in this mass region has recently been observed in the GW230529 event by the LIGO-Virgo-KAGRA collaboration. Our results naturally agree with this detection, which the most popular prescriptions for explodability and remnant masses are not able to reproduce. In general, we find that the explosion outcome and mass of the remnant strongly depend on the pre-collapse structure of the progenitor. However, their dependence on the initial mass of the star and the mass of the CO core is highly uncertain and non-linear.
\end{abstract}

\keywords{supernovae -- supernova remnants -- neutron stars -- black holes}


\maketitle

\section{Introduction}
\label{sec:intro}
Core-collapse supernovae (CCSNe) are an extremely complex phenomenon, for which more and more sophisticated models have been developed in the last several decades \citep{Colgate_White1966,Arnett1966,Bethe_Wilson1985,Bruenn1985,Miller1993_first2D,Herant1994_first2D,Janka1996_1st_parametric_3D,Fryer2002}. The last ten years have produced a large number of self-consistent exploding simulations in three dimensions from many different groups \citep{Muller2012_2D_GR,Takiwaki2012_original_3DnSNe,Lentz2015_3D,Janka2016_success_expl,Bruenn2016_expl_en,Takiwaki2016_3DnSNe_3D_explosions,OConnor2018_2D_M1,Muller2019_3Dcoconut,Burrows2020_3DFornax,Bugli2021_MHD_3D_SN,Nakamura2022_3DnSNe_binary_star_1987a}. This has significantly improved our understanding of the relevant physical processes that occur during the post-bounce and explosion phases. In particular, of crucial importance for the explosion are the Standing Accretion Shock Instability (SASI) \citep{Foglizzo2006_SASI,Fernandez2015_3D_SASI,Hanke2012_SASI_crit_lum,Marek2009_SASI_diag_ene} and, to an ever greater extent, neutrino-driven turbulent convection \citep{Couch2015_turbulence,Radice2016,Radice2018_turbulence,Murphy2013_turb_in_CCSNe}. However, multi-dimensional simulations are still too computationally expensive to explore the parameter space of supernovae (SNe) efficiently. 

One of the most important problems in CCSNe concerns the explodability as a function of Zero Age Main Sequence (ZAMS) mass, as well as the resulting explosion properties such as remnant masses, nucleosynthesis yields, light curves, explosion energies etc\dots. The only viable options to investigate a wide range of ZAMS masses and metallicities are currently spherically symmetric simulations (but see \cite{Wang2022_prog_study_ram_pressure} for a recent large suite of 2D models), semi-analytical prescriptions, or a combination of both. 

The goal of this paper is to study the explodability and remnant mass distribution of a wide range of progenitors with different ZAMS masses and metallicities. 

Neutron stars (NSs) and black holes (BHs) are the two possible remnants left behind after the successful (or failed) explosion of a supernova. Studying their formation and population can help understand several phenomena, ranging from gamma-ray bursts (GRBs), binary mergers, exotic accretion scenarios, and many others. It is therefore important to have reliable models that can predict the birth-mass distribution of NSs and BHs. 

However, mass and radius measurements of NSs are quite challenging. The vast majority of existing measurements are performed on millisecond pulsars, which are believed to have experienced some degree of mass accretion from a companion and are therefore not representative of the birth-mass distribution. Another challenge that theoretical predictions have to face is that most stellar models simulate isolated single stars, whereas most massive stars experience significant binary interactions throughout their lives.

The case of BHs is even more uncertain, both from a theoretical and observational standpoint. The best constraints on their mass are given by the measurements of the LIGO-Virgo-KAGRA (LVK) collaboration \citep{Abbott2018_GW_detectors_future_plans}. However, the low end of their mass distribution, which can be extremely useful to constrain which massive stars give birth to black holes, is affected by large uncertainties. There are a few objects that could fall in this low-mass range between 2 $M_\odot$ and 3 $M_\odot$, and classifying them as high-mass NSs or low-mass BHs is quite challenging, not to mention the uncertainties regarding their formation channel. However, a 2.5-4.5 $M_\odot$ object, which is most likely the first stellar-mass black hole ever observed, has recently been detected by the LVK collaboration \citep{LIGO2024_GW230529}, and this is the first step to putting more stringent constraints on the low-end of black hole masses. Predicting the final mass of a black hole is also a theoretical challenge, since even in failed SN events weak shocks can form, and eject a fraction of the envelope \citep{Lovegrove2013_failed_SN_fej,Lovegrove2017_failedSN_fej_sims,Fernandez2018_failedSN_fej_progs,Ivanov2021_failed_SN_fej_EOS_nu,Schneider_2023_failed_SN_fej,Antoni2023_failed_SNeII}. However, the precise mechanisms responsible for this, and the amount of matter ejected, are still uncertain.

In the past, there have been several studies aimed at studying the remnant mass distributions, based on different 1D simulations as well as semi-analytical models \citep{Timmes1996_NS_BH_birth_mass,Fryer2012_remnant_popsynth,Raithel2018_remnant_mass,Meskhi2022_EOS_remnant_mass_distr,Mandel2020_recipes_remn_kick,Fryer2022_nu_conv_remnant_masses}. For the birth-mass distribution of NSs, all of these studies predict a large peak at $\sim 1.2-1.4$ $M_\odot$, whose existence has been shown by countless works and observations. More interesting is instead the presence (or absence) of another small peak at $\sim 1.7-1.8$ $M_\odot$, which is where most of the disagreement among the previously mentioned studies rises. The discrepancy concerning the birth-mass distribution of BHs is even larger, and different studies use different prescriptions to decide what fraction of the star will contribute to the final mass of the black hole. 

In this paper, we present a simple recipe to determine the explodability and compute NS and BH masses based on the density and entropy profiles of the pre-SN progenitor.

In Section \ref{sec:explosion_problem} we present an overview of some of the previously derived explosion models, and then we briefly describe the method used in this work to simulate the explosion of 341 pre-SN models. These simulations are an extension to a larger ZAMS mass range and lower metallicity of the previously simulated models of \cite{Boccioli2023_explodability} (hereafter BR23). In Section \ref{sec:criterion} we also generalize the explodability criterion already derived by BR23 in light of the larger set of simulations performed in the present study. The main results of this paper are presented in Section \ref{sec:remnant}, where we provide theoretical mass distributions of NSs and BHs, and compare them to the most recent observations. Finally, in Section \ref{sec:conclusions} we summarize the main findings of this paper and comment on future research directions.

\section{The explosion model}
\label{sec:explosion_problem}
Simulating the explosion of CCSNe is an impressively tough feat. Only in the last couple of decades have simulations started showing reliable, robust explosions. However, these typically occur in high-fidelity multidimensional simulations that require millions of CPU hours. To date, only one study employing 100 2D simulations has been performed \citep{Wang2022_prog_study_ram_pressure}, due to their very high computational cost. The overwhelming majority of studies are performed using either semi-analytical models or spherically symmetric, 1D simulations, which are still extremely useful and widely used.

\subsection{Previous models}
The simplest way to achieve an explosion is to use the so-called piston \citep{Woosley1995_pistons} or bomb models \citep{Blinnikov1993_bomb}. The former use a moving inner boundary at some mass coordinate M$_{\rm piston}$, mimicking the expansion of the shock. The latter inject a certain amount of thermal energy at some mass coordinate M$_{\rm bomb}$ for a few seconds, which revives the shock and launches the explosion. In both cases the models are calibrated to reproduce a certain explosion energy and the amount of Nichel ejected, based on observational constraints.

Later models employed more sophisticated numerical setups, but based on the same core idea. The most well-known example is the model of \cite{Ugliano2012}, where a hydrodynamic simulation collapses the star, and then the inner core is manually excised and replaced with a contracting inner boundary emitting some neutrino luminosity $L_{\nu,c}$ according to a simplified, semi-analytical proto-neutron star (PNS) cooling model. The time evolution of the inner boundary is chosen to reproduce explosion energy and Nichel mass of SN 1987A \citep{Sonneborn1987a}. This is the model used by \cite{Ertl2016_explodability}, \cite{Sukhbold2016_explodability}, and later \cite{Raithel2018_remnant_mass}, who derived explodabilities and remnant masses for a number of stars at solar metallicity.

Another popular semi-analytical model is the one by \cite{Fryer2012_remnant_popsynth}, who assume the explosion to happen in three steps: collapse and bounce, convective engine, and eventual post-explosion fallback. The second phase is particularly important and is the fundamental reason why self-consistent spherically symmetric simulations do not show any explosions (except for very rare cases of low-mass, low-compactness stars) whereas their multi-dimensional counterparts do \citep{Herant1994_first2D,Radice2016,Radice2018_contraints_on_EOS_from_GW170817,Couch2015_turbulence}. Therefore, \cite{Fryer2012_remnant_popsynth} use a semi-analytical prescription for the convective velocity (later revised in \citep{Fryer2022_nu_conv_remnant_masses}) that adds the necessary energy to achieve an explosion. A similar approach has also been developed by \cite{Muller2016_prog_connection} and \cite{Mandel2020_recipes_remn_kick}.

Another method adopted in the past is to simulate the collapse and post-bounce phases with self-consistent, state-of-the-art spherically symmetric simulations. As mentioned above, however, simulations in spherical symmetry do not show any explosions. In these simulations, the energy deposited by neutrinos in the gain region (i.e. the region with positive net neutrino heating) is not enough to successfully launch an explosion, since convection is not present. To overcome this, one option is to artificially increase the neutrino heating in the gain region by some factor $f_{\rm heat}$. This was, for example, the method adopted by \cite{OConnor2011_explodability}, who also employed a simple prescription to model neutrino emission and absorption. They noticed that progenitors with larger compactness needed a larger value of $f_{\rm heat}$ to explode, and therefore concluded that progenitors above a certain compactness $\xi_{2.5}$ would not explode, where:
\begin{equation}
    \label{eq:compactness}
    \xi_{\rm M} = \dfrac{M/M_\odot}{R(M_{\rm bary} = M)/1000\, {\rm km}},
\end{equation}
and $R(M_{\rm bary} = M)$ is the radial coordinate that encloses a mass $M$. \cite{OConnor2011_explodability} used $\xi_{\rm 2.5}$ at the time of bounce. In the remainder of the paper, however, we will refer to the pre-SN compactness values, i.e. at the very beginning of the collapse phase, unless stated otherwise.

Another option, which is a bit more sophisticated, is to take some energy from the heavy-lepton neutrinos, which would otherwise not interact with matter outside of the PNS, and deposit it into the gain region. \cite{Perego2015_PUSH1} devised this method and applied it to simulations with a bit more sophisticated neutrino transport. They calibrated their energy deposition in such a way that progenitors with larger compactness experience less energy deposition, which was done specifically to prevent high-compactness progenitors from exploding.

The fundamental idea of all of the models described above is the same: the explosion must be triggered by depositing energy behind the shock. The first piston and bomb models simply added this energy by hand into the model. The model from \cite{Ugliano2012} was devised to take this energy from a physical source, which in their case was neutrinos. The models of \cite{Fryer2012_remnant_popsynth} and \cite{Muller2016_Oxburning} have the advantage of being relatively simple and include convective energy, which would otherwise be absent from 1D models. However, they heavily rely on semi-analytical prescriptions. The models of \cite{OConnor2011_explodability} and \cite{Perego2015_PUSH1} also take this extra energy from neutrinos, but they follow the evolution of the PNS self-consistently and, in the case of \cite{Perego2015_PUSH1}, employ a robust treatment of neutrino transport \citep{Liebendorfer2009_IDSA}. In conclusion, all of these different methods introduce the extra energy into their model using a prescription that is: (i) extremely simple but derived from a physical mechanism like convection \citep{Fryer2012_remnant_popsynth,Muller2016_Oxburning}: (ii) more sophisticated but the energy is arbitrarily transferred from neutrinos to matter without following an accurate physical model \citep{OConnor2011_explodability,Perego2015_PUSH1,Ugliano2012}. 

\subsection{1D+ model including neutrino-driven convection}
\label{sec:STIR_description}
In this paper, we will take a different approach, and use self-consistent simulations that include the effects of convection through STIR, a parametric model for convection developed by \cite{Couch2020_STIR}. The advantage of this model is that it is derived from the Reynolds decomposition of the Euler equations, where the closure adopted is chosen using a mixing-length theory (MLT) approach. Therefore, the extra energy in the model is calculated starting from a physically consistent model. It should be noted that the form of the ensuing equations is not explicitly energy-conserving \citep{Muller2019_STIR}. However, this is a minor drawback considering that, in all of the models described above, a certain amount of energy has to be somehow injected (in some cases crudely) into the model. Moreover, one could argue that STIR still effectively conserves energy if one accounts for the free energy associated with the convectively unstable thermodynamic gradients \citep{Mabanta2019_turbulence_in_CCSN,Muller2019_STIR}. The equation describing the evolution of the total energy can be derived by combining equations 27 and 29 from \cite{Couch2020_STIR}:
\begin{equation}
\begin{split}
\label{eq:total_energy}
\pder{\rho e_{\rm tot}}{ t} &+ \frac{1}{r^2}\pder{}{r} [r^2 v_r(\rho e_{\rm tot} + P + P_{\rm turb}) - r^2\rho D \nabla e_{\rm tot}] \\ 
&= -\rho v_r g + Q_\nu -\rho v_{\rm turb}^2 \pder{v_r}{r} + \rho v_{\rm turb} \omega_{\rm BV}^2 \Lambda_{\rm mix}.
\end{split}
\end{equation}
where $e_{\rm tot} = e + v_{\rm turb}^2$, $P_{\rm turb} = \rho v_{\rm turb}^2$,  $v_{\rm turb}$ is the turbulent (i.e. convective) velocity, and $D$ is a diffusion coefficient. Notice that $e$ includes the contributions from both internal and kinetic energy. The most important quantities in the model are the mixing length $\Lambda_{\rm mix}$ , and the \BV/ frequency $\omega_{\rm BV}^2$:
\begin{align}
    \label{eq:lambda_mix}
    \Lambda_{\rm mix} &= \alpha_{\rm MLT} \frac{P}{\rho g},\\
    \label{eq:BV_eff}
    \omega^2_{\rm BV} &=  g \left(\frac{1}{\rho} \pder{\rho(1 + \epsilon)}{r} - \frac{ 1}{ \rho c_s^2} \pder{P}{r} \right),
\end{align}
where $P$ is the pressure, $\rho$ is the density, and $g$ is the local gravitational acceleration. 
Notice that all of the above equations assume Newtonian gravity for simplicity. However, the model used in this paper includes general relativistic effects, which can be very important, especially for $\omega_{\rm BV}$. More details on these equations and how they were derived can be found in \cite{Boccioli2021_STIR_GR}, \cite{Mabanta2018_MLT_turb}, and \cite{Couch2020_STIR}.

The simulations presented in this paper were all run with \texttt{GR1D}\footnote{The code is publicly available at \url{https://github.com/evanoconnor/GR1D}} \citep{Boccioli2021_STIR_GR,OConnor2010,OConnor2015}. Since the original code was modified with the addition of neutrino-driven convention, an inherently multidimensional effect, we will refer to the code and the resulting 1D simulations as \texttt{GR1D+} and \texttt{1D+}, respectively.

Convection is more efficient for larger values of $\alpha_{\rm MLT}$, and therefore the shock is revived to larger radii. The procedure for calibrating $\alpha_{\rm MLT}$ is described at length in \cite{Boccioli2021_STIR_GR,Boccioli2022_EOS_effect,Boccioli2023_explodability}. To summarize it, one compares the \texttt{GR1D+} simulations to 3D simulations of the same initial progenitor. More specifically, one can compare the time evolution of the shock radius and, more importantly, the energy generated by neutrino-driven convection in the gain region. Then one can select the value of $\alpha_{\rm MLT}$ for which these quantities best match the 3D results. This value lies in the best-fit range $1.5 \leqslant \alpha_{\rm MLT} \leqslant 1.52$. As discussed in BR23, we select a value of $\alpha_{\rm MLT} = 1.51$, which we will use for the remainder of this paper.

As shown by \cite{Boccioli2021_STIR_GR}, the calibration is somewhat dependent on spatial and neutrino energy resolution. Therefore, to avoid re-calibrating, we utilize the same numerical setup used in BR23. For this study, we simulated 3 sets of pre-SN progenitors: 
\begin{enumerate}
    \item the ``Zero Metallicity'' set: 30 progenitors at zero metallicity from \cite{Woosley2002_KEPLER_models} in the mass range $11\ M_\odot \leqslant M \leqslant 40\ M_\odot$;
    \item the ``Low Metallicity'' set: 111 progenitors at a metallicity $z=10^{-4} z_\odot$ from \cite{Woosley2002_KEPLER_models} in the mass range $11\ M_\odot \leqslant M \leqslant 75\ M_\odot$;
    \item the ``Solar Metallicity'' set: 200 progenitors at solar metallicity $z= z_\odot$ from \cite{Sukhbold2016_explodability} in the mass range $9\ M_\odot \leqslant M \leqslant 120\ M_\odot$.
\end{enumerate}
For the simulations at solar metallicity, we used the existing simulations from BR23 in the mass range $12\ M_\odot \leqslant M \leqslant 28\ M_\odot$. All of the simulations presented in this paper were run for more than 2 seconds in case of explosion, long after the shock has left the computational domain. For the non-exploding ones, we only ran the simulation far enough to observe the shock fallback. All simulations were run using 18 neutrino energy groups, and adopting the SFHo equation of state (EOS) for nuclear matter at high density \citep{Hempel2010_HS_RMF,Steiner2013_SFHo}. More details on the simulation setup are given in BR23.

\section{Explodability of massive stars}
\label{sec:criterion}
The problem of predicting the explosion outcome of supernovae based on pre-SN properties is of great theoretical and practical interest. From a theoretical point of view, it can inform as to what the cause of the explosion is, and provide a causal connection between pre-SN properties and the explosion dynamics. From a practical point of view, knowing which stars explode and which ones do not can help guide population synthesis and galactic chemical evolution simulations in determining explosion outcome and remnant properties of a star based on its thermodynamic structure.

Most population synthesis codes \citep{Iorio2023_SEVN_popsynth,Riley2022_COMPAS_popsynth,Breivik2020_COSMIC_popsynth,Eldridge2017_BPASS_popsynth,Izzard2004_binaryc_popsynth,Kamlah2022_BSE-LEVELC_popsynth,Mennekens2014_Brussels_popsynth,Belczynski2008_Stratrack_popsynth}, however, only carry information about the Carbon-Oxygen (CO) core of massive stars. As shown by recent population studies of core-collapse supernovae, the explosion outcome is sensitive to the structure of the Silicon and Oxygen shell formed after the end of silicon burning. The dependence of the pre-collapse thermodynamic profiles on the ZAMS mass of the star and the mass of the CO core is affected by several uncertainties, varies among different codes \citep{Chieffi2020_presupernova_models}, and can be highly non-linear. Therefore, reliable predictions for the explodability in population synthesis codes can only be done if the structure of the pre-collapse star is computed. This is currently beyond most codes' capabilities, but promising efforts towards having access to the full thermodynamic structure of the star are underway \cite{Fragos2023_POSYDON_methods_paper}.

In this paper, we provide remnant masses and explosion outcome predictions based on the final pre-SN structure of the star. These predictions could eventually be expressed in terms of properties of the CO core using a procedure similar to what was done in \cite{Patton2020_popsynth_prescription}, but that goes beyond the scope of this paper and is left for future work.

\subsection{The explosion criterion of BR23}
\begin{figure*}
\includegraphics[width=\textwidth]{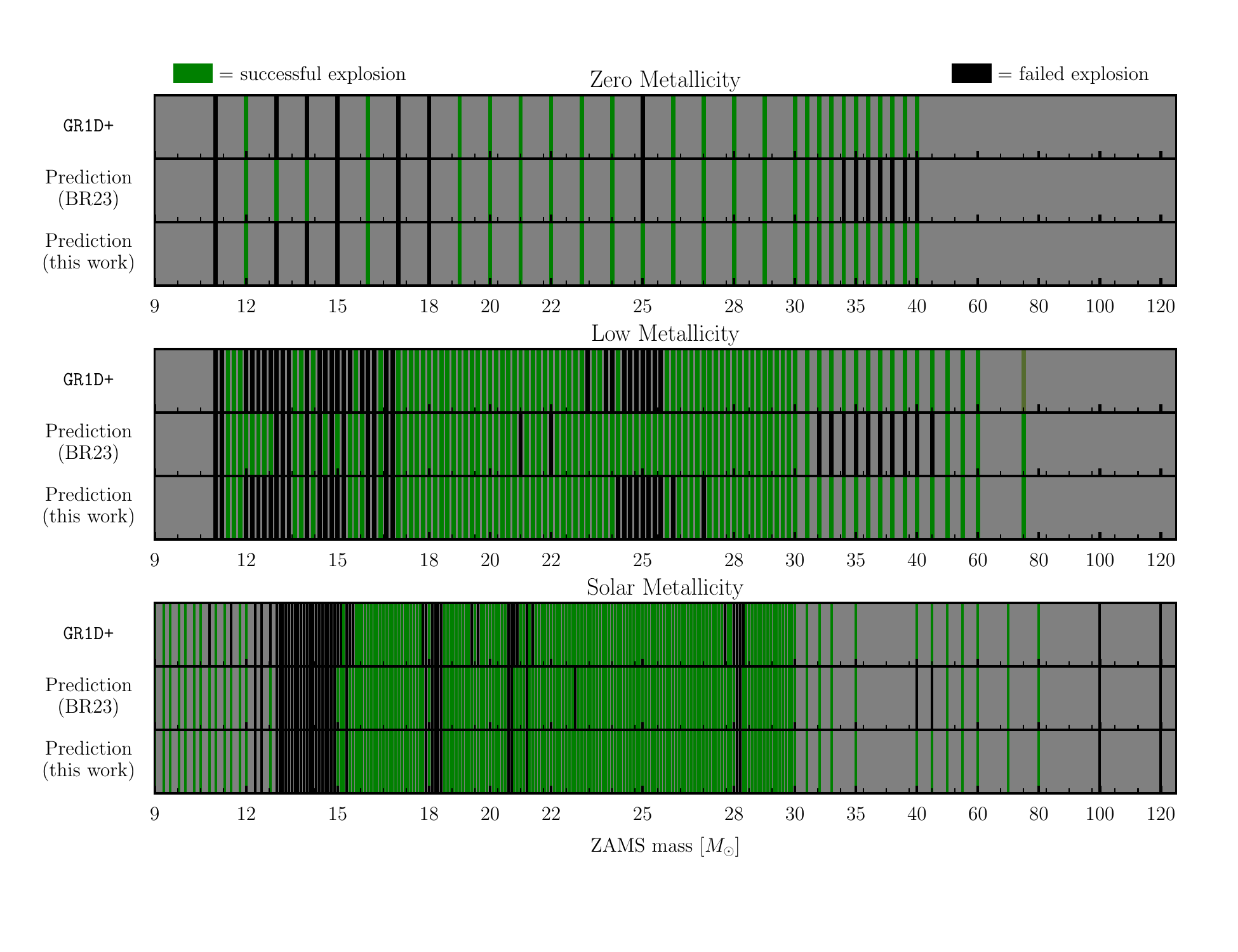}
\caption{\label{fig:expl_pattern}The upper, middle, and lower panels show the outcome of the 1D+ simulations for zero, low, and solar metallicity simulations described in the text. The low metallicity refers to $z=10^{-4}\ z_\odot$. In this work, a successful shock revival is considered a successful explosion. The first line of each panel shows the outcome of the simulations carried out with \texttt{GR1D+}. The second line shows the predicted outcome according to the criterion of BR23. The third line refers to the predicted outcome according to the criterion described in section \ref{sec:criterion}. Grey areas indicate that no pre-SN progenitor at that ZAMS mass was available, and therefore no simulation was performed.}
\end{figure*}

The explodability of massive stars has been a highly debated topic in the past few years. Historically, it was believed that stars less massive than $\sim 20$ $M_\odot$ would explode in a supernova and form a neutron star. Stars more massive than $\sim 20$ $M_\odot$ would instead lead to failed supernovae and form a black hole. Recently, this has been challenged by several studies \citep{Ertl2016_explodability,Sukhbold2016_explodability,Muller2016_prog_connection,Couch2020_STIR,Boccioli2023_explodability,Wang2022_prog_study_ram_pressure}. It is now well accepted that the general picture is quite different. The explodability as a function of Zero Age Main Sequence (ZAMS) mass is non-monotonic. Therefore, there are ``islands of explodability'' in certain mass ranges, and failed SN in others. The explodability as a function of ZAMS mass according to our simulations is shown in the upper panels of Figure \ref{fig:expl_pattern}. For example, for the solar metallicity set there is an island of failed SNe between 12 $M_\odot$ and 15 $M_\odot$, around 18 $M_\odot$, 21 $M_\odot$, 28 $M_\odot$, and above 100 $M_\odot$. The location of these ``islands of explodability" is however a debated topic. For example, \cite{Ertl2016_explodability} derived an explosion criterion using simulations where the explosion was artificially triggered. Using their method for the same solar metallicity set presented here, they showed that most stars explode except for the mass ranges $22\ M_\odot \lesssim M \lesssim 25\ M_\odot$ and $28\ M_\odot \lesssim M \lesssim 50\ M_\odot$. 

Other studies, which employed either 2D simulations \citep{Wang2022_prog_study_ram_pressure} or 1D+ simulations like the ones presented here \citep{Boccioli2023_explodability,Couch2020_STIR}, showed instead that stars in the mass ranges $22\ M_\odot \lesssim M \lesssim 25\ M_\odot$ and $28\ M_\odot \lesssim M \lesssim 50\ M_\odot$ explode, whereas stars in the mass range $12\ M_\odot \lesssim M \lesssim 15\ M_\odot$ yield failed explosions. The results of all three studies are in excellent agreement with each other and, more importantly, with a suite of a dozen 3D simulations from \citep{Burrows2020_3DFornax}. In those 3D simulations, the range $22\ M_\odot \lesssim M \lesssim 25\ M_\odot$ was not covered (only one 25 $M_\odot$ star, from a slightly different progenitor set, was simulated), but no stars in the range $12\ M_\odot \lesssim M \lesssim 15\ M_\odot$ yielded explosions, whereas all of the others did, in agreement with the three studies mentioned above. 

An important caveat to this discussion is that the ZAMS mass ranges quoted here are only valid for the specific sets of stellar evolution calculations used to produce the pre-SN progenitors. As shown in BR23, the stellar evolution code (and the assumptions and approximations used in the stellar evolution calculations) can modify the explodability as a function of ZAMS mass quite significantly \citep{Boccioli2023_explodability,Boccioli2024_Review}. Therefore, one should be cautious when referring to the CCSN outcome of specific ZAMS masses and metallicities, since the evolution of those stars is quite uncertain. As we will show in the remainder of this paper, what matters for the outcome of a CCSN are the thermodynamic profiles of that star at the end of its life. Therefore, one can say with (relative) certainty whether a specific pre-SN density profile will lead to an explosion. However, what initial ZAMS mass and metallicity produce that density profile is more ambiguous, due to the large uncertainties that affect the numerical evolution of massive stars.

In BR23, an explodability criterion was derived which predicts whether a given star explodes based on the thermodynamic structure of the pre-SN progenitor. Here, we briefly summarize their criterion, and then illustrate how we extended it in light of the wider range of ZAMS masses range analyzed in this paper and, more importantly, in light of the two additional progenitors sets at zero and low metallicity. The BR23 criterion states that:
\begin{enumerate}
    \item if $\widetilde{t}_{\rm accr}$ > 0.4 s, the star will not explode;
    \item if $\widetilde{t}_{\rm accr}$ < 0.4 s, the star will explode if $\delta \rho_{\rm Si/O}^2/\rho_{\rm Si/O}^2 > 0.08$.
\end{enumerate}
Here, $\rho_{\rm Si/O}$ is the density of the pre-SN progenitor at which the Si/O interface is located (see Section 5.2 of BR23), $\delta \rho_{\rm Si/O}$ is the magnitude of the density drop at that interface, and $\widetilde{t}_{\rm accr}$ is the time after bounce when the Si/O interface is accreted through the shock:
\begin{equation}
    \label{eq:taccr}
    \widetilde{t}_{\rm accr} = C t_{\rm ff} - t_0 = C \sqrt{ \frac{\pi}{4 G \bar{\rho}}} - t_0~~.
\end{equation}
The coefficients $C$ and $t_0$ were derived by fitting this equation to the actual time of accretion of the Si/Si-O interface found in the simulations. The values found by BR23 for KEPLER progenitors were $C = 0.78$ and $t_0 = 0.13$ s. Since the progenitors at zero and low metallicity of \cite{Woosley2002_KEPLER_models} were also computed with the KEPLER code, we don't expect these coefficients to significantly change once we add these progenitors to the fit. As expected, once we add the new simulations performed at lower metallicities and the ones at solar metallicity for $M < 12$ $M_\odot$ and $M > 28$ $M_\odot$, we obtain the same value of $t_0 = 0.13$ s and a slightly lower value of $C = 0.76$. For consistency with previous work, we chose to adopt $C = 0.78$, but we verified that both values of $C$ lead to the same results.

The Si/O interface is particularly relevant for the explosion because the entropy jump, and therefore the density drop, is very pronounced, which causes the ram pressure of the infalling material to decrease significantly. It is also important to point out that, even though we refer to it as a Si/O interface for simplicity, in most cases, the largest entropy jump (i.e. density drop) is located inside the silicon shell, where a pocket of oxygen has formed (e.g. see left panel of Figure 4 from BR23). Therefore, a more appropriate definition would be the Si/Si-O interface. For conciseness purposes, and since it has largely been referred to as Si/O interface in the SN literature, we adopt the latter in all subscripts hereafter. However, it is important to note that this could be a misleading naming convention.

The addition of progenitors at low and zero metallicities (in particular the ones with masses $M > 30$ $M_\odot$) required a slight change in the explodability criterion. Those progenitors, compared to the ones analyzed in BR23, have very large compactnesses and are more likely to explode. This suggests that a more nuanced criterion, which also considers compactness, should be used. A detailed analysis of the role of compactness is currently underway \citep{Boccioli2024_comp_vs_Qdot}, but for the purposes of this article, a simple phenomenological criterion can instead be used:

\begin{enumerate}
    \item if $\xi_{2.0}$ > 0.5, the star will explode;
    \item if $\xi_{2.0}$ < 0.5 and $\widetilde{t}_{\rm accr}$ > 0.3 s, the star will not explode;
    \item if $\xi_{2.0}$ < 0.5 and $\widetilde{t}_{\rm accr}$ < 0.3 s, the star will explode if $\delta \rho_{\rm Si/O}^2/\rho_{\rm Si/O}^2 > 0.08$.
\end{enumerate}
This simple criterion significantly improves the prediction rate compared to the one by BR23, as shown in Figure \ref{fig:expl_pattern} and Table \ref{tab:criteria_performance}. It also suggests that compactness might play an important role in determining the explosion.

Finally, it should be noted that the explosion outcome of high-compactness progenitors is still a topic of debate. Traditionally, high-compactness, very high-mass stars were considered to lead to failed SN and form black holes, as discussed in Section \ref{sec:explosion_problem}. Even 3D simulations have yet to agree on whether a supernova can be produced by these high-compactness progenitors, and what the remnant of these events will be \citep{Chan2018_BH_SN_40Msol,Burrows2023_BH_supernova_40Msol}. It should be stressed that the simulations presented in this paper are conducted in spherical symmetry, and therefore no fallback is present after the explosion sets in. All of the successful explosions leave behind a neutron star, with only one exception described in Section \ref{sec:remnant_NS}. However, 3D simulations of high compactness progenitors \citep{Chan2018_BH_SN_40Msol,Burrows2023_BH_supernova_40Msol}, show that, in some cases, despite an explosion develops, late-time fallback can accrete enough mass on the central object to produce a small black hole. The amount of late-time fallback is highly uncertain since long-term simulations that follow the shock until it breaks out of the envelope are computationally extremely challenging. Only a few of these simulations have been performed \citep{Wongwathanarat2015_3D_shk_breakout,Muller2018_stripped_SN_to_breakout,Stockinger2020_3D_low_mass_to_breakout,Sandoval2021_3D_shk_breakout}, and all of them are for relatively low mass (up to 20 $M_\odot$), low-compactness progenitors. Only more simplistic, semi-analytical prescriptions have been used so far to model late-time fallback to investigate whether small-mass BHs can be produced in the explosion of high-mass, high-compactness progenitors, both in 1D \citep{Liu2021_remnant_LMG} and 3D \citep{Vigna-Gomez2021_fallback_Gadget}.

\begin{table}
\centering 
\begin{ruledtabular}
\begin{tabular}{>{\centering\arraybackslash}p{15mm}|c|c|c}
{} & \shortstack{Zero\\Metallicity} & \shortstack{Low\\Metallicity} & \shortstack{Solar\\Metallicity} \\
\midrule
{BR23} & \shortstack{\centering} & \shortstack{} & \shortstack{} \\
\midrule
FP &                          6.7\% &                        17.1\% &                           8.0\% \\
FN &                         23.3\% &                        10.8\% &                           1.5\% \\
\midrule
{This work} & \shortstack{} & \shortstack{} & \shortstack{} \\
\midrule
FP &                          3.3\% &                         4.5\% &                           8.0\% \\
FN &                          0.0\% &                         2.7\% &                           0.0\% \\
\end{tabular}
\end{ruledtabular}

\caption{False positives (FP) and false negatives (FN) produced by the criterion of BR23 versus the one derived in this work when compared to the results of the simulations for the zero, low, and solar metallicity sets described in section \ref{sec:STIR_description}. \label{tab:criteria_performance}}
\end{table}

\section{Remnant masses from simulations}
\label{sec:remnant}
In this section, we analyze the properties of the CCSN remnants obtained from simulations. Moreover, we present a recipe to predict the masses of both NSs and BHs, based on the pre-SN properties of the progenitor stars.

\subsection{Neutron Stars}
\label{sec:remnant_NS}
As mentioned above, our simulations were run in spherical symmetry, and therefore no fallback is present. Once the explosion sets in, the accretion stops, and the baryonic mass of the PNS will remain constant until the end of the simulation. By solving the TOV equation for the given EOS, it is straightforward to convert the baryonic mass into the gravitational mass of the cold NS. This is the NS mass M$_{\rm NS}$ that we will refer to hereafter. 

All of the simulations that successfully produce an explosion leave behind a NS, with the only exception of the 75 $M_\odot$ progenitor from the low metallicity set. This particular progenitor has the largest compactness ($\xi_{2.0} = 0.99$) of all progenitors and shows a successful shock revival occurs at around 0.2 s. However, due to its large compactness, its PNS baryonic mass is very large, i.e. $M_{\rm PNS} = 2.57$ $M_\odot$. This causes the simulation to stop at $\sim 0.4$ s, since the PNS is collapsing to a BH. Indeed, a baryonic mass of 2.57 $M_\odot$ would correspond to a gravitational mass of roughly 2.13 $M_\odot$, which is larger than the maximum mass of a cold NS allowed by the SFHo EOS, which is $\sim 2.05$ $M_\odot$. We decided to exclude this progenitor from our analysis since it is unclear whether the initial shock expansion would continue after BH formation. Neutrino emission will shut off, and this has often been shown to cause the shock to fall back without being able to break out of the star. On the other hand, a weak shock might still be launched \citep{Lovegrove2013_failed_SN_fej,Lovegrove2017_failedSN_fej_sims,Fernandez2018_failedSN_fej_progs,Ivanov2021_failed_SN_fej_EOS_nu,Schneider_2023_failed_SN_fej}, which would unbind a larger fraction of the envelope compared to cases where BH formation occurs without any shock revival at all (discussed in the next Section). Rather than handling the fallback for this one progenitor \textit{ad hoc}, we prefer excluding it from the analysis altogether for consistency. Moreover, this one 75 $M_\odot$ progenitor would be greatly disfavored by the initial mass function (IMF), and therefore excluding it will not appreciably affect our results and conclusions.

\begin{figure}
\includegraphics[width=\columnwidth]{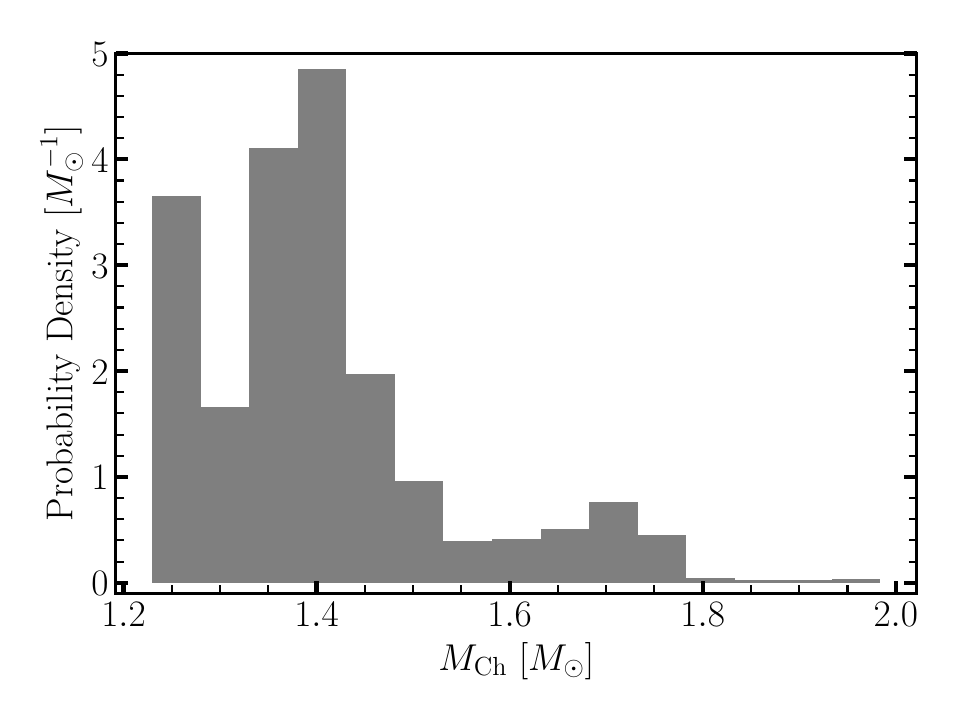} \\
\caption{\label{fig:CH_mass} Chandrasekhar mass calculated using eq. \eqref{eq:M_ch} for all metallicities combined, weighted by an IMF with an exponent of $-2.35$.}
\end{figure} 

Generally speaking, the more compact the star (i.e. the shallower the density profile is), the larger the mass of the neutron star will be since the mass accretion during the stalling phase will be larger. This intuitively suggests that the neutron star's mass is tightly correlated with the compactness of the star. The latter is (mostly) the consequence of convective phenomena occurring during the post-main sequence evolution. Intuitively, one can also expect the remnant mass to be tightly correlated with the mass of the iron core. More specifically, the mass of the iron core is a lower bound for the final mass of the remnant, since the Fe-peak elements present inside the core cannot be ejected. If they were, we would observe a much higher abundance of such elements \citep{Weaver1978_presn_evolution}, which we do not. A more accurate predictor of the final remnant mass could however be the Chandrasekhar mass, which is generally considered to be $\sim 1.4 M_\odot$. However, as described in \cite{Timmes1996_NS_BH_birth_mass}, this value can vary depending on the electron fraction and entropy in the core:
\begin{equation}
    \label{eq:M_ch}
        M_{\rm Ch} = M_{\rm Ch,0} \left[1 + \left(\dfrac{s_e}{\pi Y_e}\right)^2\right] = 5.38 Y_e^2 \left[1 + \left(\dfrac{s_e}{\pi Y_e}\right)^2\right]  ,
\end{equation}
where $s_e$ and $Y_e$ are the electron entropy per baryon and electron fraction in the core, respectively. Typically, the Chandrasekhar mass is evaluated at the pre-collapse stage, which we consider to be the time when the velocity drops below 1000 km/s inside the core. It can be shown that the entropy in the core $s_e$ is mostly determined by the details of Carbon burning \citep{Sukhbold2020_missing_red_supergiant}. Above a certain mass, central C-burning occurs in radiative rather than convective equilibrium, which changes the subsequent stages of evolution and, essentially, completely bypasses the neutrino-cooling phase \citep{Sukhbold2020_missing_red_supergiant}, which is responsible for decreasing the central entropy of the core. This eventually leads to an abrupt increase of iron core mass, central entropy, and compactness above a certain mass. As \cite{Timmes1996_NS_BH_birth_mass} pointed out, this would naturally lead to a bimodal neutron star mass distribution. This is illustrated in Figure \ref{fig:CH_mass}, and we will come back to it later in the section.

In the literature, another quantity that has been shown to correlate with the mass of the neutron star is M$_4$, i.e. the mass coordinate of the layer where the specific entropy per baryon of the pre-SN progenitor rises above 4 \citep{Ertl2016_explodability,Sukhbold2016_explodability,Patton2020_popsynth_prescription}. This layer almost always coincides with (or is very close to) the Si/Si-O interface. As showed by BR23 and \cite{Wang2022_prog_study_ram_pressure}, as well as by several 3D simulations \citep{Vartanyan2021_Binary_stars_SiO_interface,Lentz2015_3D,Muller2019_3Dcoconut,Couch2015_turbulence}, the explosion sets in relatively quickly after this layer is accreted, and therefore one expects a correlation between the mass coordinate of this layer and the final mass of the remnant. Contrary to previous studies, we do not use M$_4$, but rather we consider the exact mass location of the Si/Si-O interface M$_{\rm Si/O}$, i.e. the enclosed mass below the density $\rho_{\rm Si/O}$, since it gives a much better correlation with M$_{\rm NS}$. The location of the Si/Si-O interface is calculated according to the definition of BR23.

\begin{figure*}
\includegraphics[width=\textwidth]{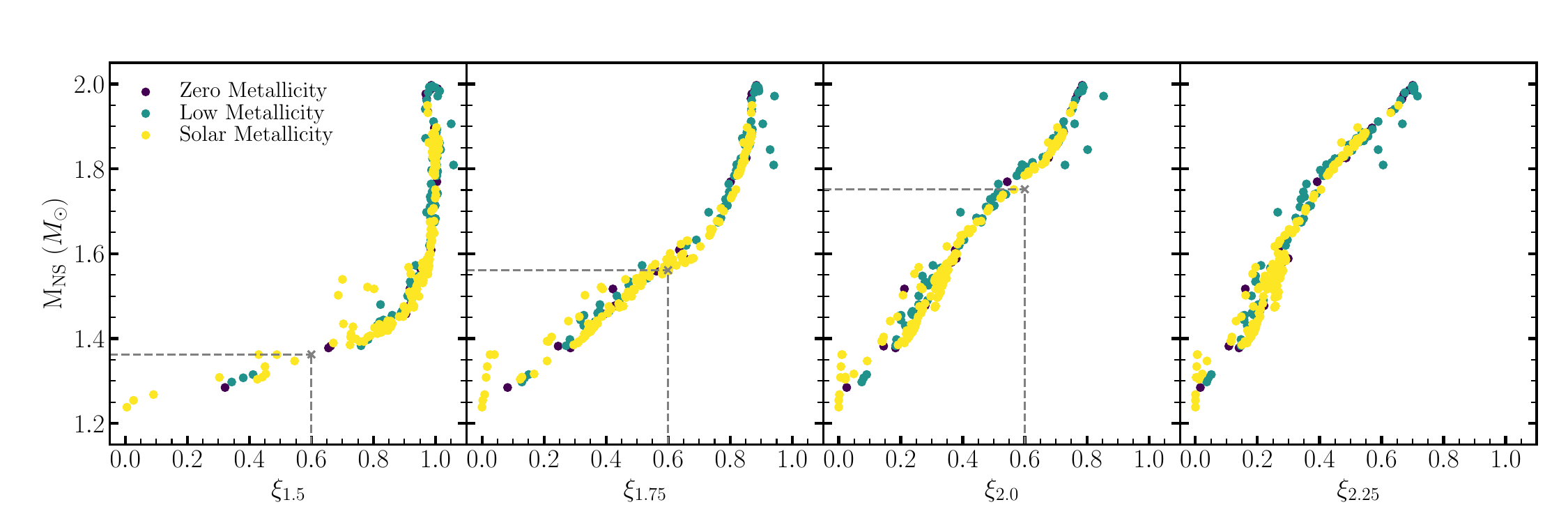}
\caption{\label{fig:Mns_vs_all_nofits}Gravitational mass of the NS as a function of compactness of the pre-SN progenitor, as defined by eq. \eqref{eq:compactness}. The grey crosses indicate the points where the dependence of M$_{\rm NS}$ on compactness steepens. Specifically, the abscissa of each cross is located at $\xi_{1.5}=0.6$, $\xi_{1.75}=0.6$, and $\xi_{2.0}=0.6$ (see the fit in Eq. \eqref{eq:comp_fit}). The ordinate of each cross is instead located at 1.36 $M_\odot$, 1.56 $M_\odot$, and 1.75 $M_\odot$, i.e. the gravitational mass corresponding to a baryonic mass of 1.5 $M_\odot$, 1.75 $M_\odot$, and 2.0 $M_\odot$, respectively. The fact that the crosses lie along the data points shows that our choice of $\xi_{1.5}=0.6$, $\xi_{1.75}=0.6$, and $\xi_{2.0}=0.6$ is consistent with the idea that Nss with baryonic masses above 1.5 $M_\odot$, 1.75 $M_\odot$, and 2.0 $M_\odot$, cannot be described using $\xi_{1.5}$, $\xi_{1.75}$, and $\xi_{2.0}$, which is why the curve steepens above that threshold.}
\end{figure*} 

\begin{figure*}
\includegraphics[width=\textwidth]{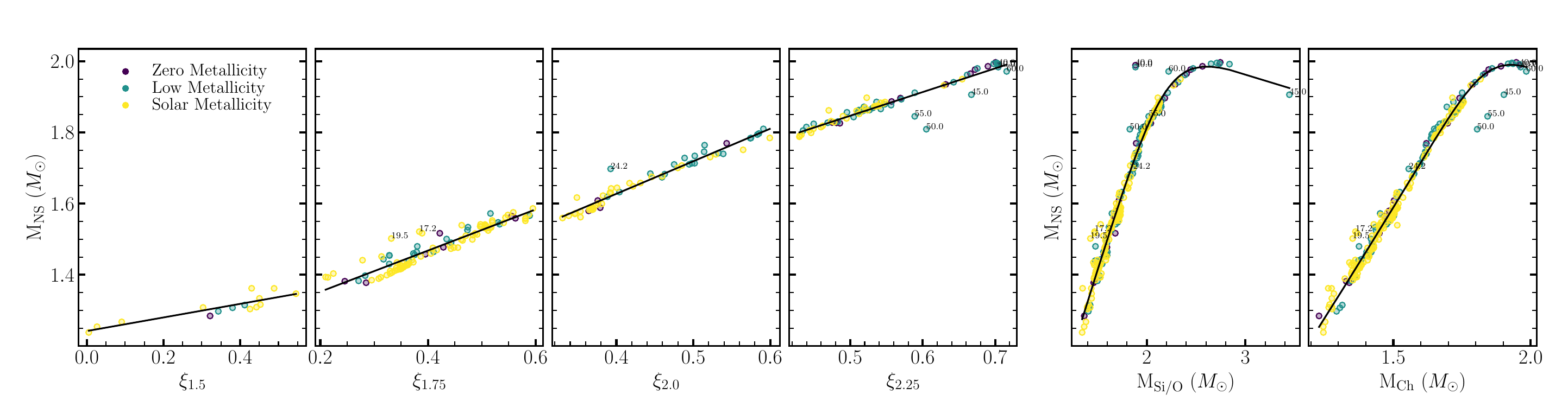}
\caption{\label{fig:all_fits} The four leftmost panels show the correlation between different compactnesses and the gravitational mass of the NS. Each dot represents a different simulation for progenitors with zero (yellow), low (green), and zero (purple) metallicities. The black solid lines show the fit described in eq. \eqref{eq:comp_fit}. The penultimate (last) panel shows the correlation between $M_{\rm Si/O}$ ($M_{\rm Ch}$) and the gravitational mass of the NS The ZAMS mass. The solid lines in the two rightmost panels represent the broken power-law fits (eq. \eqref{eq:broken_pl} described in the text. The ZAMS mass of the "outliers" is also shown as an annotation.}
\end{figure*}

This correlation is shown in the first of the two rightmost panels of Figure \ref{fig:all_fits}. It is worth pointing out that the outliers all lie in the upper left region of the plots, because the deviation from the trend is caused by the fact that the explosion for those progenitors sets in much later compared to when the Si/Si-O interface is accreted through the shock, and therefore more mass is accreted onto the central PNS. As expected from the discussion above, the NS mass $M_{\rm NS}$ is also highly dependent on the Chandrasekhar mass $M_{\rm Ch}$ of the pre-supernova star. This is shown in the second of the two rightmost panels of Figure \ref{fig:all_fits}. The correlation between $M_{\rm NS}$ and $M_{\rm Ch}$ is quite good, although some clear outliers can be seen at low metallicity (the 50, 55, and 60 $M_\odot$ progenitors), as well as a few at solar metallicity for $M_{\rm Ch} < 1.4 M_\odot$.

The tightest correlation can be seen between M$_{\rm NS}$ and compactness calculated at different mass coordinates: $\xi_{1.5}$, $\xi_{1.75}$, $\xi_{2.0}$, and $\xi_{2.25}$ (left 4 panels of Figure \ref{fig:all_fits}). The choice to pick these four different compactnesses can be understood by analyzing Figure \ref{fig:Mns_vs_all_nofits}. The correlation between M$_{\rm NS}$ and $\xi_{1.5}$ is very good at low values of $\xi_{1.5}$, and then gets progressively worse at larger values. This is not surprising since, for NSs with M$_{\rm NS} \gtrsim 1.36$ $M_\odot$ (corresponding to a baryonic mass of 1.5 $M_\odot$), what determines the mass of the remnant is not only the accretion of the layers within an enclosed mass of 1.5 $M_\odot$, but also the accretion of layers outside that mass shell. Therefore, even if the density profile up to an enclosed mass of 1.5 $M_\odot$ is very similar, if the density outside this layer is different then the resulting mass of the remnant will be different. This explains why M$_{\rm NS}$ diverges as a function of $\xi_{1.5}$ once M$_{\rm NS}$ goes above $\sim 1.36$ $M_\odot$. A similar argument holds for $\xi_{1.75}$ and $\xi_{2.0}$, for which the steepening of the trend occurs at values of M$_{\rm NS}$ above $\sim 1.56$ $M_\odot$ and $\sim 1.75$ $M_\odot$, respectively, and it is not as pronounced as it is for $\xi_{1.5}$. Finally, if one looks at M$_{\rm NS}$ as a function of $\xi_{2.25}$, the trend is linear at large masses, without any steepening. This is again not surprising since a baryonic mass of 2.25 $M_\odot$ corresponds to a NS with a gravitational mass of roughly 1.94 $M_\odot$, which is very close to the maximum NS mass of 2.05 $M_\odot$. Therefore, only the density profile inside an enclosed mass of 2.25 $M_\odot$ dictates the final mass of the NS.

It is also interesting to notice that there is a clear correlation between M$_{\rm NS}$ and $\xi_{1.5}$ at small values of compactness. However, for $\xi_{1.75}$, $\xi_{2.0}$, and $\xi_{2.25}$, the stars with the smallest compactness show essentially a divergence in M$_{\rm NS}$. The reason is very similar to the divergence (or steepening) that happens at large values of compactness. For these low-compactness stars, the final NS mass is dictated by the accretion history of the inner $\sim 1.2-1.5$ $M_\odot$. However, these stars have very shallow density profiles, and therefore very extended envelopes, which makes their compactnesses $\xi_{1.75}$, $\xi_{2.0}$, and $\xi_{2.25}$ extremely small. Their interior structure could however be quite different, as demonstrated by their different values of $\xi_{1.5}$, which explains why they have very different M$_{\rm NS}$ despite having very similar (practically zero) $\xi_{1.75}$, $\xi_{2.0}$, and $\xi_{2.25}$. This effect can also be seen, to a less dramatic extent, in the steepening of M$_{\rm NS}$ as a function of $\xi_{2.25}$ for compactnesses below $\sim 0.4$.

To summarize, there is a clear correlation between $M_{\rm Si/O}$ and M$_{\rm NS}$, and an even better one between compactness and M$_{\rm NS}$. In particular, the final NS mass seems to be best predicted by the compactness calculated at an enclosed mass that is not too much smaller, nor larger, than the final NS mass. Therefore, we find that a very simple fit, which is at the same time very accurate, is the following:
\begin{equation}
    \label{eq:comp_fit}
    M_{\rm NS} = 
    \begin{cases}
    \begin{aligned}
        0.191 & \times \xi_{1.5} + 1.242& &\xi_{1.5} < 0.6~; \\
        0.578 & \times \xi_{1.75} + 1.237& &\xi_{1.5} \geq 0.6~,\ \xi_{1.75} < 0.6~; \\
        0.914 & \times \xi_{2.0} + 1.262& &\xi_{1.5} \geq 0.6~,\ \xi_{1.75} \geq 0.6~,\\
        &&& \xi_{2.0} < 0.6~; \\
        0.665 & \times \xi_{2.25} + 1.514& &\xi_{2.0} \geq 0.6~. \\
    \end{aligned}
    \end{cases}
\end{equation}
This is shown in the left 4 panels of Figure \ref{fig:all_fits}, and will be referred to as $\xi$-fit hereafter.

Similarly, we fit a broken power-law to M$_{\rm NS}$ as a function of M$_{\rm Si/O}$ using the python package \texttt{Astropy} \citep{Astropy2022_v5.0}. The functional form is the following:
\begin{equation}
    \label{eq:broken_pl}
    f(x) = A \left( \frac{x}{x_b} \right) ^ {-\alpha_1}
       \left\{
          \frac{1}{2}
          \left[
            1 + \left( \frac{x}{x_b}\right)^{1 / \Delta}
          \right]
       \right\}^{(\alpha_1 - \alpha_2) \Delta}~,
\end{equation}
where x=M$_{\rm Si/O}$ and $f(x)=$ M$_{\rm NS}$. This will be referred to as $M_{\rm Si/O}$-fit hereafter. The best-fit values for the parameters are $A=1.934$ $M_\odot$, $x_b=2.245$ $M_\odot$, $\alpha_1=-0.938$, $\alpha_2=0.163$, and $\Delta=0.085$.

The same functional form with x=M$_{\rm Ch}$ is used to fit M$_{\rm NS}$ as a function of the Chandrasekhar mass. This will be referred to as $M_{\rm Ch}$-fit hereafter. The best-fit values for the parameters are $A=1.928$ $M_\odot$, $x_b=1.790$ $M_\odot$, $\alpha_1=-1.194$, $\alpha_2=-0.163$, and $\Delta=0.027$.

It is interesting to note that the outliers are roughly the same progenitors for all of the fits, although as can be seen from Figure \ref{fig:all_fits}, that is not always the case. Due to the large differences compared to the bulk of all of the other progenitors, the 45, 50, and 55 $M_\odot$ progenitors of the low metallicity set have been excluded from the M$_{\rm Ch}$-fit. Indeed, their M$_{\rm NS}$ is equal to their M$_{\rm Ch}$, whereas for all of the other progenitors the NS mass is always larger than M$_{\rm Ch}$. Because of the steep dependence of M$_{\rm NS}$ on M$_{\rm Si/O}$ and M$_{\rm Ch}$, the NS masses for progenitors similar to the 17.2 $M_\odot$ and 19.5 $M_\odot$ progenitors are significantly underestimated by the fit. However, the dependence of M$_{\rm NS}$ on the different compactnesses is not as steep, and therefore the underestimation is not as dramatic. This will be shown more clearly in the next section.

\subsection{Theoretical Neutron Star Mass Distributions}
\label{sec:remnant_NS_distr}
The observed population of Neutron Star is known to have a large peak around $\sim 1.4$ $M_\odot$, and there is strong evidence for a second peak (or at least an extended tail) at around $\sim 1.9$ $M_\odot$, as shown by \citep{Antoniadis2016_NS_mass_distr}. This is based on the observation of hundreds of millisecond pulsars found in our Galaxy. Some of these are known to have experienced accretion from a companion star, which increased their mass compared to when they were formed. Therefore, one should compare the stars simulated in this work only to single-star systems or binary systems with little to no mass transfer \citep{Meskhi2022_EOS_remnant_mass_distr}. One category of objects that satisfies this requirement is slow-spinning pulsars, whose low spin indicates very little mass accretion \citep{Ozel2016_NS_obs_review}. Another category of objects is double neutron star (DNS) systems, where two neutron stars orbit each other. In this case, however, it is less clear how close they are to their birth masses \citep{Ozel2012_birth_mass_NS,Kiziltan2013_NS_mass_distr}, since some of these systems are expected to have experienced some degree of accretion. At the same time, they tend to have very low masses, as well as a very narrow mass distribution (as explained below), which suggests a very clear evolutionary path.

\begin{figure*}
\includegraphics[width=\textwidth]{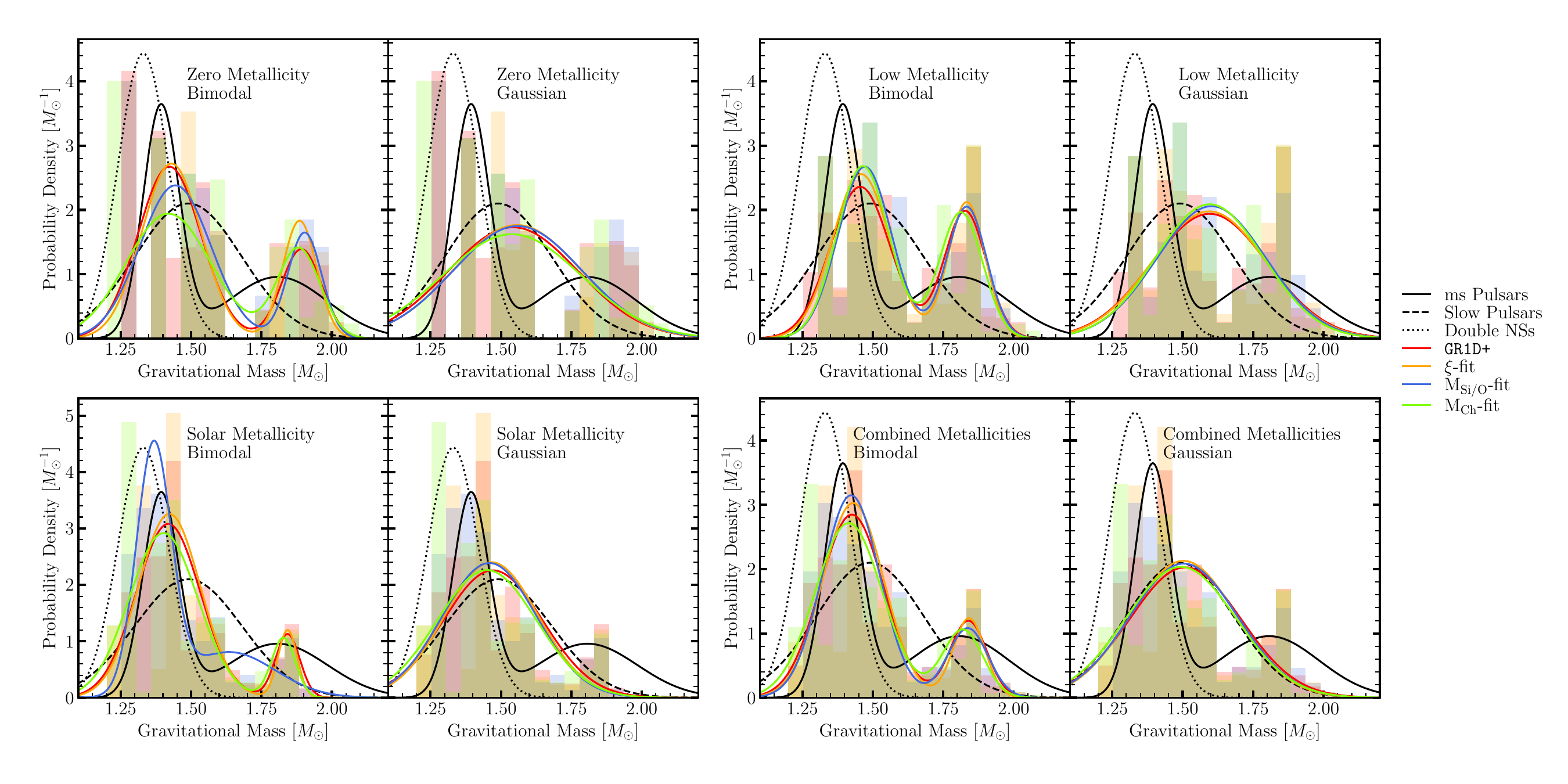}
\caption{\label{fig:NS_distr} Comparison between theoretical (in colors) and observed (in black) gravitational mass distributions of NSs. The shaded regions show the histograms of the raw data, and the solid colored lines show the fitted distributions for NS masses calculated from simulations (in red, set (ii) in section \ref{sec:remnant_NS_distr}), from the $\xi$-fit (in orange, set (ii) in section \ref{sec:remnant_NS_distr}), from the $M_{\rm Si/O}$-fit (in blue, set (iii) in section \ref{sec:remnant_NS_distr}), and from the $M_{\rm Ch}$-fit (in green, set (iv) in section \ref{sec:remnant_NS_distr}). The solid black line shows the distribution of millisecond pulsars from \cite{Antoniadis2016_NS_mass_distr}, and the dashed and dotted black lines show the distribution of double NSs and slow pulsars, respectively, from \cite{Ozel2016_NS_obs_review}. }
\end{figure*}

This leaves us with three observed NS mass distributions. The probability density function (PDF) of slow-spinning pulsars and double neutron stars is described by a simple Gaussian:
\begin{equation}
    \label{eq:Gaussian}
    P(M_{\rm NS}) = \frac{1}{\sigma\sqrt{2\pi}} \exp\left[- \frac{(M_{\rm NS}-\mu)^2}{2\sigma}\right],
\end{equation}
where $\mu = 1.49$ $M_\odot$ and $\sigma = 0.19$  $M_\odot$ for slow pulsars (dashed black line in Figure \ref{fig:NS_distr}) and $\mu = 1.33$ $M_\odot$ and $\sigma = 0.09$  $M_\odot$ for double neutron stars (dotted black line in Figure \ref{fig:NS_distr}). The bimodal distribution of millisecond pulsars from \cite{Antoniadis2016_NS_mass_distr} (solid black line in Figure \ref{fig:NS_distr}) is a combination of two Gaussians in the form of Eq. \eqref{eq:Gaussian} weighted by a relative ratio $r$:
\begin{equation}
    \label{eq:antoniadis}
    P(M_{\rm NS}) = (1-r)\ {\rm G}(\mu_1, \sigma_1) + r\ {\rm G}(\mu_2, \sigma_2),
\end{equation}
where $\mu_1 = 1.393$ $M_\odot$, $\sigma_1 = 0.064$ $M_\odot$, $\mu_2 = 1.807$ $M_\odot$, $\sigma_2 = 0.177$ $M_\odot$, and $r = 0.425$. The distribution of slow-spinning NS has been derived for a relatively small population of only $\sim 12$ objects, whereas the population of DNS has roughly twice as many objects and the population of millisecond pulsars is quite large, counting $> 100$ objects.

To compute mass distributions from the simulations, we assume a Salpeter IMF (initial mass function) with an exponent $\alpha = -2.35$ \citep{Salpeter1955_IMF}, independent of metallicity. The NS masses are then calculated by converting the baryonic mass of the PNS at the end of the simulation to the corresponding gravitational mass of a cold NS by solving the TOV equation. This provides an empirical NS mass distribution directly from the simulations. Moreover, one can establish which stars explode as predicted by our explodability criterion described in Section \ref{sec:criterion}, and then calculate the NS mass based on the $\xi$-fit, M$_{\rm Si/O}$-fit, and M$_{\rm Ch}$-fit described in this Section. Therefore, this leads to four sets of NS masses: (i) the one calculated directly from \texttt{GR1D+} simulations; (ii) the one predicted by our explodability criterion and the $\xi$-fit; (iii) the one predicted by our explodability criterion and the M$_{\rm Si/O}$-fit; (iv) the one predicted by our explodability criterion and the M$_{\rm Ch}$-fit. The advantage of calculating NS masses from (ii), (iii), and (iv) is that they only depend on the density, entropy, and electron fraction profiles of the pre-SN progenitor, without the need to perform any CCSN simulations. These four sets of NS masses can then be used to compute NS mass distributions (see Figure \ref{fig:NS_distr}) by weighting them with the IMF.

\begin{table}
\centering 
\begin{ruledtabular}
\begin{tabular}{>{\centering\arraybackslash}p{12.5mm}|cccc}

\multicolumn{5}{c}{Gaussian Fit} \\
\midrule
\midrule
{$\mu$} & \shortstack{Zero} & \shortstack{Low} & \shortstack{Solar} & \shortstack{All} \\
\midrule
\texttt{GR1D+}   &                           1.54 &                          1.59 &                            1.47 &                                 1.51 \\
$\xi$-fit        &                           1.56 &                          1.59 &                            1.47 &                                 1.51 \\
M$_{\rm Si/O}$-fit        &                           1.56 &                          1.61 &                            1.46 &                                 1.50 \\
M$_{\rm Ch}$-fit &                           1.54 &                          1.60 &                            1.45 &                                 1.49 \\
\midrule
{$\sigma$} & \shortstack{Zero} & \shortstack{Low} & \shortstack{Solar} & \shortstack{All} \\
\midrule
\texttt{GR1D+}   &                           0.23 &                          0.21 &                            0.18 &                                 0.20 \\
$\xi$-fit        &                           0.23 &                          0.20 &                            0.17 &                                 0.19 \\
M$_{\rm Si/O}$-fit        &                           0.23 &                          0.19 &                            0.17 &                                 0.19 \\
M$_{\rm Ch}$-fit &                           0.25 &                          0.19 &                            0.18 &                                 0.20 \\
\midrule
\midrule
\multicolumn{5}{c}{Bimodal Fit} \\
\midrule
\midrule
{$\mu_1$} & \shortstack{Zero} & \shortstack{Low} & \shortstack{Solar} & \shortstack{All} \\
\midrule
\texttt{GR1D+}   &                           1.42 &                          1.45 &                            1.42 &                                 1.43 \\
$\xi$-fit        &                           1.43 &                          1.46 &                            1.42 &                                 1.43 \\
M$_{\rm Si/O}$-fit        &                           1.44 &                          1.47 &                            1.37 &                                 1.42 \\
M$_{\rm Ch}$-fit &                           1.42 &                          1.46 &                            1.36 &                                 1.41 \\
\midrule
{$\sigma_1$} & \shortstack{Zero} & \shortstack{Low} & \shortstack{Solar} & \shortstack{All} \\
\midrule
\texttt{GR1D+}   &                           0.11 &                          0.11 &                            0.11 &                                 0.11 \\
$\xi$-fit        &                           0.11 &                          0.10 &                            0.11 &                                 0.11 \\
M$_{\rm Si/O}$-fit        &                           0.12 &                         0.094 &                           0.062 &                                 0.10 \\
M$_{\rm Ch}$-fit &                           0.15 &                         0.092 &                           0.089 &                                 0.12 \\
\midrule
{$\mu_2$} & \shortstack{Zero} & \shortstack{Low} & \shortstack{Solar} & \shortstack{All} \\
\midrule
\texttt{GR1D+}   &                           1.89 &                          1.83 &                            1.84 &                                 1.84 \\
$\xi$-fit        &                           1.88 &                          1.83 &                            1.84 &                                 1.84 \\
M$_{\rm Si/O}$-fit        &                           1.90 &                          1.83 &                            1.64 &                                 1.83 \\
M$_{\rm Ch}$-fit &                           1.88 &                          1.81 &                            1.66 &                                 1.82 \\
\midrule
{$\sigma_2$} & \shortstack{Zero} & \shortstack{Low} & \shortstack{Solar} & \shortstack{All} \\
\midrule
\texttt{GR1D+}   &                          0.071 &                         0.073 &                           0.041 &                                0.062 \\
$\xi$-fit        &                          0.062 &                         0.066 &                           0.032 &                                0.057 \\
M$_{\rm Si/O}$-fit        &                          0.060 &                         0.071 &                            0.16 &                                0.072 \\
M$_{\rm Ch}$-fit &                          0.077 &                         0.078 &                            0.15 &                                0.071 \\
\midrule
{$r$} & \shortstack{Zero} & \shortstack{Low} & \shortstack{Solar} & \shortstack{All} \\
\midrule
\texttt{GR1D+}   &                           0.25 &                          0.37 &                            0.12 &                                 0.20 \\
$\xi$-fit        &                           0.27 &                          0.36 &                            0.10 &                                 0.19 \\
M$_{\rm Si/O}$-fit        &                           0.25 &                          0.36 &                            0.33 &                                 0.19 \\
M$_{\rm Ch}$-fit &                           0.27 &                          0.37 &                            0.29 &                                 0.19 \\

\end{tabular}
\end{ruledtabular}

\caption{Best-fit parameters for the Gaussian and bimodal fits for the four sets of theoretical NS masses described in section \ref{sec:remnant_NS_distr} \label{tab:NS_best_fit}}
\end{table}

From observations we expect the NS mass distribution to be either Gaussian (Eq. \eqref{eq:Gaussian}) or bimodal (Eq. \eqref{eq:antoniadis}). Therefore, we fit both distributions to the data and then compare them to determine which one best fits the data. All the fits were performed with the "fit" method from the python module \texttt{statsmodels} \citep{statsmodels2010_py_package}. To perform the fit, we generated 20,000 synthetic samples based on the theoretical NS masses weighted with the IMF, for each of the four sets of NS masses described above. The best-fit bimodal and Gaussian distributions are shown in Figure \ref{fig:NS_distr}, and the best-fit parameters are summarized in Table \ref{tab:NS_best_fit}. Finally, in Table \ref{tab:ML_KS_NS} we show the value of the maximum log-likelihood, as well as the value of the KS test and the corresponding p-value for all of the fits. As can be easily concluded by looking at the histograms, the bimodal distribution fits the data much better, as confirmed by the much larger log-likelihood, for all metallicities. 

We also performed a KS test to investigate whether the fitted distribution is compatible with the data. The detailed procedure is reported in Appendix \ref{sec:appendix_NS_distr}. With the null hypothesis being that the data was drawn from the fitted distribution, a p-value above 0.05 indicates that the null hypothesis cannot be rejected. A value below 0.05 indicates that the null hypothesis can be rejected with 95 \% confidence. Unsurprisingly, the bimodal fit always has a p-value > 0.05, with only a few exceptions. More interestingly, the Gaussian distribution is compatible with the data for the zero and low metallicity sets, whereas it is not for solar metallicities and for all metallicities combined. 

However, one should carefully interpret these results. As explained in appendix \ref{sec:appendix_NS_distr}, the result of the KS test generally depends on the number of samples used for the test, and the more one generates, the smaller the result of the KS test. In other words, the p-value depends on the number of bins chosen when binning the data. The fewer bins one has, the higher the p-value will be, since with less information one cannot constrain the empirical distribution. The KS test results reported in Table \ref{tab:ML_KS_NS} were obtained by choosing a number of bins that is half of the total number of data points. The dependence of the KS test on the number of bins is shown in Appendix \ref{sec:appendix_NS_distr}.

This naturally brings us to compare our theoretical results with the observed distributions. We will focus on the comparison between the observed distributions and the theoretical result for all metallicities combined. Firstly, the $\xi$-fit, the M$_{\rm Si/O}$-fit, and the M$_{\rm Ch}$-fit reproduce the distribution from the \texttt{GR1D+} simulations quite well, with the $\xi$-fit being better at predicting the variance of the two peaks, which is not surprising considering the outliers at solar metallicities seen in the M$_{\rm Si/O}$-fit and M$_{\rm Ch}$-fit in Figure \ref{fig:all_fits}, which are responsible for widening the distribution (see also the third panel of Figure \ref{fig:NS_distr}). All four theoretical distributions have peaks at the same locations, which are systematically slightly larger than the peak locations of the distribution from \cite{Antoniadis2016_NS_mass_distr}, but qualitatively very similar.

Usually, as mentioned above, a possible explanation of the bimodal nature (or extended tail) of the observed population of ms pulsars is that the NS birth-mass distribution is a Gaussian centered at around 1.4 $M_\odot$. Then, some NSs will accrete matter via binary interactions, hence giving rise to the second peak (or extended tail) seen by \cite{Antoniadis2016_NS_mass_distr}. However, nothing prevents the birth-mass distribution itself from being bimodal (\"{O}zel, private communication), as seen in our simulations. Moreover, a bimodal birth-mass distribution means that producing pulsars with large masses does not necessarily require accreting a significant amount of mass. 

The birth-mass distribution we find has a wider low-mass peak and a narrower high-mass peak compared to \citep{Antoniadis2016_NS_mass_distr}. In this scenario, both low-mass and high-mass NSs could accrete mass via binary interactions, which could give rise to a widening of the peak at high masses and, potentially, also to a narrowing of the peak at low masses.

It is also instructive to discuss the Gaussian fit to the distribution even though a KS test can already conclude that our theoretical NSs cannot be drawn from Gaussian, as discussed above. As can be seen in the last panel of Figure \ref{fig:NS_distr}, the Gaussian fits for all four theoretical distributions are in extremely good agreement with the distribution of slow pulsars. As mentioned above, this distribution has been derived from a small sample of pulsars, and therefore a bimodal nature of the underlying distribution cannot be excluded. Moreover, the agreement between the theoretical Gaussian fits and the observed distribution suggests that, if the underlying distribution was indeed bimodal, a Gaussian fit would produce exactly the distribution that has been observed. In other words, if we assume that the true birth-mass distribution is equal to the bimodal distribution we find in the simulations, a small sample of objects would be indeed distributed according to the observed distribution of slow pulsars. We want to stress that this is a speculative conclusion since no robust statistical argument supports it. Nonetheless, it shows that robust conclusions about the bimodality (or lack thereof) of the birth-mass distribution of NSs cannot be made without larger samples of observed slow pulsars. 

As can be seen by looking at the distribution of double NSs, our results seem to be underproducing low-mass NSs. This could have a few different explanations. Firstly, the narrow peak of the observed distribution suggests a preferred evolution channel of these stars, and therefore it should be most likely combined with other observations (like slow pulsars, for example), to properly derive a birth-mass distribution, which could shift the peak to larger masses. Secondly, our simulated progenitors are stars that developed an iron core with masses of 9 $M_\odot$ and above in the solar metallicity set, and 11 $M_\odot$ and above in the zero and low metallicity sets. However, electron-capture supernovae at around 8-10 $M_\odot$ are responsible for producing the lowest mass NSs and will be very much favored by the IMF. Therefore, the theoretical distributions derived here likely exclude a large number of low-mass NSs which could explain the population represented by double NSs. Moreover, this would be compatible with the fact that the vast majority of binaries are composed by two relatively small massive stars (find reference).

As mentioned above, when combining all of the metallicities, the $\xi$-fit, the M$_{\rm Si/O}$-fit, and the M$_{\rm Ch}$-fit correctly reproduce the results from the simulations. However, unsurprisingly, at solar metallicity the M$_{\rm Si/O}$-fit and M$_{\rm Ch}$-fit grossly misrepresents the distribution obtained from simulations. This could have been intuitively concluded by noticing the large deviation in the NS masses predicted by both fits compared to the results of the simulations for some solar-metallicity progenitors (see Figure \ref{fig:all_fits}). Therefore, we recommend using the $\xi$-fit rather than one based on M$_{\rm Si/O}$ or M$_{\rm Ch}$ since it is less susceptible to progenitors exploding long after the accretion of the Si/Si-O interface. As discussed in the previous section, the stars whose M$_{\rm NS}$ is poorly reproduced by the $M_{\rm Si/O}$-fit and $M_{\rm Ch}$-fit, explode much later than the accretion of the Si/Si-O interface, and therefore both fits systematically underestimate the mass of the final compact object. A detailed study of what causes the explosion of these stars to be so different is beyond the scope of this paper and is currently underway \citep{Boccioli2024_FEC+_SiO}.

To summarize, our simulations show a clear bimodal distribution across all metallicities. The bimodal nature of the Chandrasekhar mass distribution shown in Figure \ref{fig:CH_mass} naturally translates into a bimodal NS birth-mass distribution, as had been already recognized by \cite{Timmes1996_NS_BH_birth_mass}. The difference with previous work \citep{Ertl2016_explodability,Muller2016_prog_connection,Ebinger2020_PUSH} is that the stars with larger Chandrasekhar masses do not successfully explode according to those studies. However, with our parametric model for $\nu$-driven convection, we expect them to explode and therefore make up the second peak of the birth mass distribution. The large discrepancy in the explodability predicted by our model and the studies mentioned above is illustrated in Figure \ref{fig:N20_vs_STIR}, and discussed in more detail at the end of the section \ref{sec:remnant_BH} The comparison with observations showed that a bimodal NS birth-mass distribution would indeed be possible, and would also naturally explain the high-mass NSs without needing to accrete a large amount of mass. The best predictor for the final NS mass is the compactness of the progenitor right before collapse. Therefore, without needing to perform any CCSN simulation, using the explodability criterion described in Section \ref{sec:criterion} and the $\xi$-fit from Eq. \eqref{eq:comp_fit} one can predict the outcome of the supernova and the mass of the remnant. For the cases in which the explosion is not successful, a BH will be formed, as discussed in the next section.

\begin{table}
\centering 
\begin{ruledtabular}
\begin{tabular}{>{\centering\arraybackslash}p{12.5mm}|cccc}

\multicolumn{5}{c}{Gaussian Fit} \\
\midrule
\midrule
{MLL} & \shortstack{Zero} & \shortstack{Low} & \shortstack{Solar} & \shortstack{All} \\
\midrule
\texttt{GR1D+}   &                           1163 &                          3277 &                            6334 &                                 4098 \\
$\xi$-fit        &                           1349 &                          3594 &                            7626 &                                 4881 \\
M$_{\rm Si/O}$-fit        &                           1117 &                          4365 &                            7256 &                                 4996 \\
M$_{\rm Ch}$-fit &                           -455 &                          4628 &                            6457 &                                 4180 \\
\midrule
{KS-test} & \shortstack{Zero} & \shortstack{Low} & \shortstack{Solar} & \shortstack{All} \\
\midrule
\texttt{GR1D+}   &                           0.31 &                          0.20 &                            0.20 &                                 0.16 \\
$\xi$-fit        &                           0.30 &                          0.21 &                            0.18 &                                 0.17 \\
M$_{\rm Si/O}$-fit        &                           0.30 &                          0.20 &                            0.21 &                                 0.17 \\
M$_{\rm Ch}$-fit &                           0.28 &                          0.21 &                            0.18 &                                 0.15 \\
\midrule
\shortstack{p-value} & \shortstack{Zero} & \shortstack{Low} & \shortstack{Solar} & \shortstack{All} \\
\midrule
\texttt{GR1D+}   &                           0.17 &                         0.074 &                          0.0057 &                               0.0034 \\
$\xi$-fit        &                           0.19 &                         0.055 &                          0.0084 &                              0.00092 \\
M$_{\rm Si/O}$-fit        &                           0.19 &                         0.072 &                         0.00082 &                              0.00082 \\
M$_{\rm Ch}$-fit &                           0.26 &                         0.057 &                          0.0079 &                               0.0046 \\
\midrule
\midrule
\multicolumn{5}{c}{Bimodal Fit} \\
\midrule
\midrule
{MLL} & \shortstack{Zero} & \shortstack{Low} & \shortstack{Solar} & \shortstack{All} \\
\midrule
\texttt{GR1D+}   &                           6645 &                          7095 &                           10352 &                                 8185 \\
$\xi$-fit        &                           7920 &                          8272 &                           11889 &                                 9378 \\
M$_{\rm Si/O}$-fit        &                           6141 &                          8527 &                           12663 &                                 9188 \\
M$_{\rm Ch}$-fit &                           2384 &                          8245 &                            9309 &                                 7170 \\
\midrule
{KS-test} & \shortstack{Zero} & \shortstack{Low} & \shortstack{Solar} & \shortstack{All} \\
\midrule
\texttt{GR1D+}   &                           0.27 &                          0.16 &                            0.14 &                                 0.11 \\
$\xi$-fit        &                           0.27 &                          0.15 &                            0.14 &                                 0.12 \\
M$_{\rm Si/O}$-fit        &                           0.28 &                          0.16 &                            0.11 &                                 0.11 \\
M$_{\rm Ch}$-fit &                           0.26 &                          0.16 &                            0.18 &                                 0.12 \\
\midrule
\shortstack{p-value} & \shortstack{Zero} & \shortstack{Low} & \shortstack{Solar} & \shortstack{All} \\
\midrule
\texttt{GR1D+}   &                           0.30 &                          0.24 &                           0.082 &                                 0.12 \\
$\xi$-fit        &                           0.31 &                          0.27 &                           0.056 &                                0.041 \\
M$_{\rm Si/O}$-fit        &                           0.27 &                          0.24 &                            0.24 &                                0.071 \\
M$_{\rm Ch}$-fit &                           0.32 &                          0.24 &                          0.0086 &                                0.051 \\

\end{tabular}
\end{ruledtabular}

\caption{Properties of the best-fit distributions for the four sets of theoretical NS masses described in section \ref{sec:remnant_NS_distr}. The MLL is the Maximum Log-Likelihood of the best-fit distribution. The KS-test is the value of the Kolmogorov-Smirnov test, and the p-value is the probability of rejecting the null hypothesis that the raw data are not drawn from the fitted distribution. To perform the KS test, we binned the NS masses by choosing a number of equal-width bins equal to half of the total number of NS masses in each of the four sets of theoretical NS masses. A more detailed description of the procedure is given in appendix \ref{sec:appendix_NS_distr}. \label{tab:ML_KS_NS}}
\end{table}

\subsection{Black Holes}
\label{sec:remnant_BH}
The origin of black holes as a consequence of stellar collapse is highly uncertain. The detection of a failed SN is particularly challenging since the disappearance of a star can oftentimes be compatible with obscuration due to dust. Therefore, the same event could be interpreted as a failed SN or a sudden obscuration of the source due to a dusty environment caused by some ejection event \citep{Byrne2022_no_failed_SNe, Kochanek2023_dust_obscures_transients}. Nonetheless, it is possible to set some constraints to the fraction of failed SNe $f_{\rm fSNe}$, i.e. $f_{\rm fSNe} = 0.16^{+0.23}_{-0.12}$ according to \cite{Neustadt2021_failedSN_frac} and $f_{\rm fSNe} < 0.23$ for sources with absolute magnitudes $> -14$ according to \cite{Byrne2022_no_failed_SNe}. 

\begin{figure}
\includegraphics[width=\columnwidth]{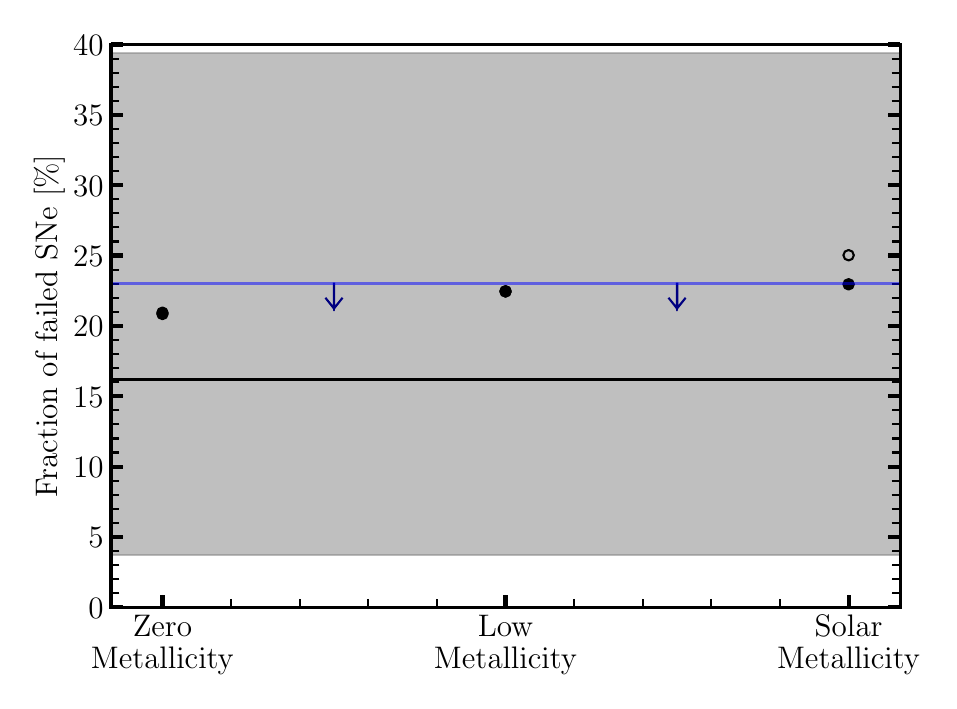}
\caption{\label{fig:failed_frac}Fraction of failed SNe for the three different sets of metallicities. The horizontal black line and grey shaded region are the fraction of failed SNe inferred by \cite{Neustadt2021_failedSN_frac} and its 90 \% confidence interval, respectively. The horizontal blue line shows the most stringent upper limit (i.e. sources with absolute magnitudes < -14) from \cite{Byrne2022_no_failed_SNe}. The filled circles show the fraction of failed SNe calculated directly from our simulations, with the assumption that stars between 8 $M_\odot$ and the smallest ZAMS mass in each set (i.e. 11 $M_\odot$ for the zero and low metallicity sets, and 9 $M_\odot$ for the solar metallicity set) successfully explode. The empty circle is instead the fraction of failed SNe assuming that the $10.75\ M_\odot$ progenitor in the solar metallicity set successfully explodes, although in our simulations it fails. }
\end{figure}

In this study, we compute the fraction of failed SNe by assuming that stars with ZAMS masses around 8 $M_\odot$ (which we did not include in our simulations due to the lack of available stellar evolution models) lead to successful supernovae, regardless of metallicity. The fate of stars around this mass, that can end their life either as a White Dwarf (WD) or as a NS, is quite uncertain. Typically, these progenitors are not very well studied \citep{Limongi2024_SAGB_ECSNe}, in part due to their complexity and the large uncertainties that affect their evolution. Moreover, simulating the neutrino-driven mechanism as well as the initial flame propagation is very challenging, although some multi-dimensional simulations have very recently started to explore this \citep{Zha2022_ECSNe_2D,Jones2016_ECSN_3D,Stockinger2020_3D_low_mass_to_breakout}. The lowest mass stars that will lead to electron-capture supernovae, and therefore produce a NS rather than a WD, are believed to be around 8-8.5 $M_\odot$ \citep{Limongi2024_SAGB_ECSNe}, and the fraction of failed SNe can vary by a few percent depending on where this lower boundary is. The fraction of failed SNe from our simulations is shown in Figure \ref{fig:failed_frac}. Notice that the fraction of failed SNe is extremely sensitive to the fate of stars on the lower end of the mass range. Therefore, for the zero and low metallicity progenitors, we assume that all progenitors between 8 $M_\odot$ and 11 $M_\odot$ explode. However, in the solar metallicity set, we found that the 10.75 $M_\odot$ progenitor does not explode, which is surprising considering that all of the low-mass progenitors should most likely explode, and it could be due to stochastic fluctuations and/or to the large uncertainties affecting stellar evolution. This alone affects the explosion fraction tremendously, as shown in Figure \ref{fig:failed_frac}. If one assumes that this progenitor explodes, and the lower end of electron-capture SNe is located at $\sim$ 8 $M_\odot$, the fraction of failed SNe found in our simulations is consistent with all current constraints.

Another uncertainty regarding stellar-mass black holes involves their birth masses, which heavily depend on the amount of fallback experienced by the central compact object. In other words, the black hole mass distribution depends on what fraction of the outer envelope will be ejected, and what fraction will be accreted onto the central compact object. The ejection of the outer envelope can be a consequence of several different mechanisms \citep{Lovegrove2013_failed_SN_fej,Lovegrove2017_failedSN_fej_sims,Fernandez2018_failedSN_fej_progs,Ivanov2021_failed_SN_fej_EOS_nu,Schneider_2023_failed_SN_fej,Antoni2023_failed_SNeII}, and the fraction of the envelope that will be ejected will likely vary among different progenitors. For simplicity, we adopt a common approach often found in the literature \citep{Patton2020_popsynth_prescription,Fryer2012_remnant_popsynth,Fryer2022_nu_conv_remnant_masses,Ebinger2020_PUSH}, where we assume that a fraction $f_{\rm ej}$ of the hydrogen envelope will be ejected, and use the same value for all progenitors.

Similarly to what was done for NSs in Section \ref{sec:remnant_NS}, we present two different sets of theoretical BH masses: (i) the one calculated directly from \texttt{GR1D+} simulations; (ii) the one predicted by the explodability criterion presented in Section \ref{sec:criterion}. Once established which stars will produce a failed SN based on (i) and (ii), the BH mass $M_{\rm BH}$ is computed by subtracting the ejected mass from the total pre-supernova mass $M_{\rm presn}$:
\begin{equation}
    \label{eq:MBH}
    M_{\rm BH} = M_{\rm presn} - f_{\rm ej} M_{\rm H},
\end{equation}
where $M_{\rm H}$ is the mass of the hydrogen envelope and $f_{\rm ej}$ is the fraction of the hydrogen envelope that has been ejected.

\begin{figure*}
\includegraphics[width=\textwidth]{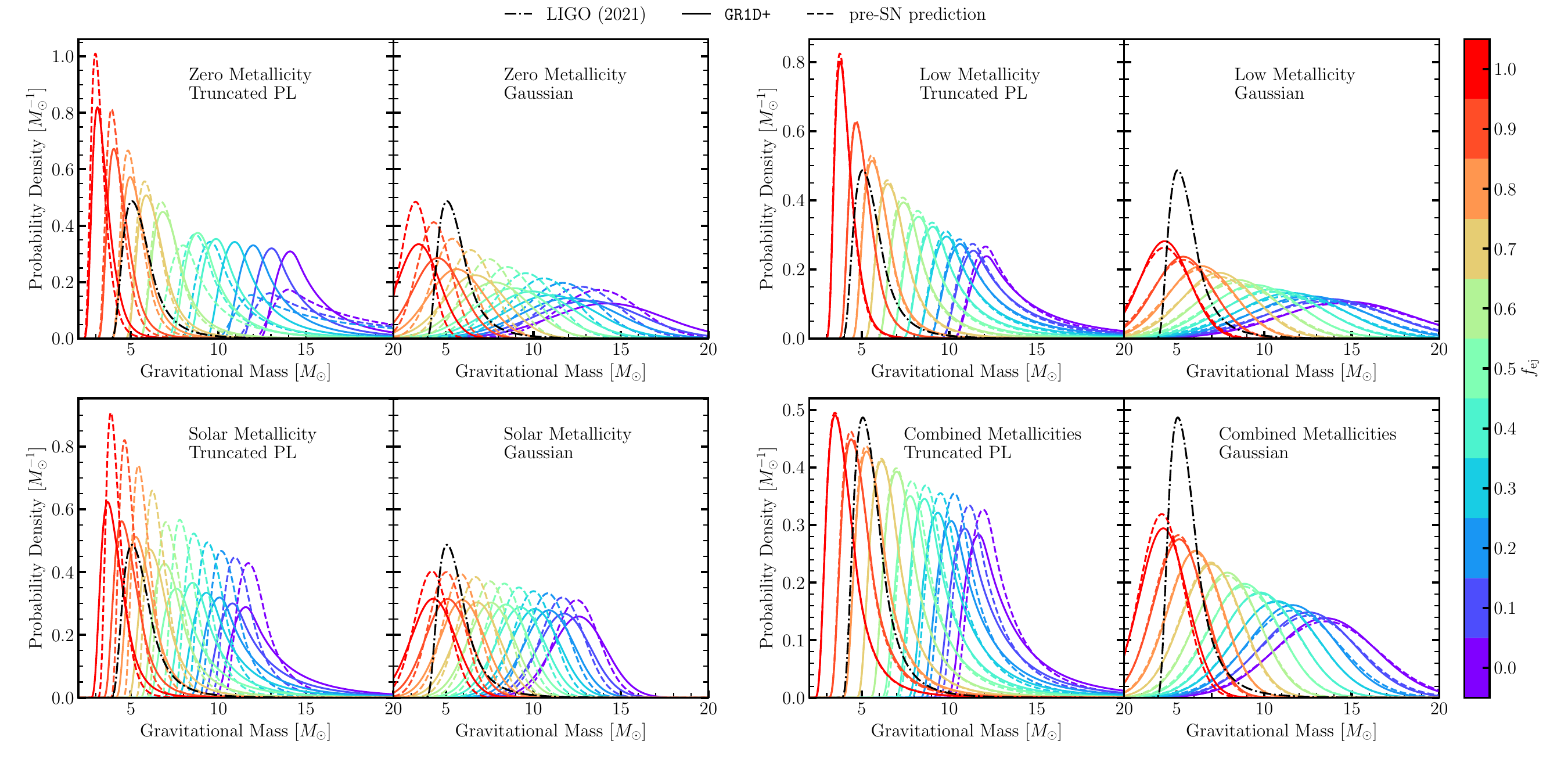}
\caption{\label{fig:BH_distr}Comparison between theoretical (in colors) and observed (in black) gravitational mass distributions of NSs. Different colors correspond to theoretical distributions calculated assuming that different fractions $f_{\rm ej}$ of the hydrogen envelope have been ejected (see eq. \eqref{eq:MBH}). Solid lines correspond to theoretical distributions derived directly from simulations (set (i) in section \ref{sec:remnant_BH}), dashed lines correspond to theoretical distributions derived using the explodability criterion from section \ref{sec:criterion} (set (i) in section \ref{sec:remnant_BH}). The dotted-dashed line shows the observed distribution inferred by the LVK collaboration, including the most recent GW230529 obsrvation \citep{LIGO2024_GW230529}. }
\end{figure*}

The LVK collaboration has significantly improved our understanding of the mass distribution of BHs. In particular, the recently announced event GW230529 completely revolutionized what we know about stellar mass black holes (i.e. BHs formed in CCSNe). The more massive companion in this binary merger event is an object with mass $3.6^{+0.8}_{-1.2} $ $M_\odot$ at $>90 \%$ confidence level, with a probability $> 90\ \%$ that this object is indeed a BH and not a very massive NS. This is a much higher confidence compared to GW190814 \citep{Abbott2020_2.6_object}, in which the 2.6 $M_\odot$ cannot be clearly classified as a high-mass NS or a low-mass BH. Therefore, one can conclude that GW230529 shows the existence of stellar black holes in the low-mass gap. Given the very recent announcement of this event and the large uncertainty in the mass of the larger object, we decided to use the truncated power law previously derived by \cite{Abbott2021_Population_properties_O2} to describe the mass distribution. However, instead of using $m_{\rm min} = 6.8 M_\odot$ as the lower end of the BH mass distribution, we use the recently-derived value of $m_{\rm min} = 3.6 M_\odot$.

The observed BH mass distribution we adopted is the following \citep{Abbott2021_Population_properties_O2} :
\begin{equation}
    \label{eq:BH_distr}
    P(M_{\rm BH}) \propto S(m | m_{\rm min},\delta_m) m^{-\alpha},
\end{equation}
where $m_{\rm min} = 3.6$ $M_\odot$, $\delta m = 3$ $M_\odot$, $\alpha = 7.1$, and $S(m | m_{\rm min},\delta_m)$ is a smoothing function defined as:
\begin{equation}
    \label{eq:BH_smoothing}
    S(m | m_{\rm min},\delta_m) =
    \begin{cases}
    \begin{aligned}
        &0 & (m<m_{\rm min})& \\
        &\frac{1}{f(m') + 1} & (m_{\rm min} \leqslant m < m_{\rm min} + \delta_m)& \\
        &1 & (m \geqslant m_{\rm min} + \delta_m)&,
    \end{aligned}
    \end{cases}
\end{equation}
with $m' = m-m_{\rm min}$ and
\begin{equation}
    \label{eq:BH_fsmooth}
    f(m') = \exp{\left[\frac{\delta_m}{m'} + \frac{\delta_m}{m' - \delta_m}\right]}.
\end{equation}
This is shown as a dotted dashed black line in Figure \ref{fig:BH_distr}. 

Similarly to what was done in the previous section, we consider two possible shapes for our simulated data: a Gaussian and a truncated power law (PL) in the form of Eq. \ref{eq:BH_distr}, where we allow both $\alpha$ and m$_{\rm min}$ to vary, and we keep $\delta_m=3$ $M_\odot$ since it is simply a smoothing parameter. For the remainder of this section, we exclude the 100 $M_\odot$ and 120 $M_\odot$ progenitors at solar metallicity from part of the analysis. These are Wolf-Rayet stars that have lost all of their hydrogen envelopes and have pre-SN masses around $\sim$ 6 $M_\odot$, much smaller than any of the other stars. Since they lost all of their hydrogen envelope, according to equation \eqref{eq:MBH} they will produce BHs of $\sim 6 M_\odot$, regardless of $f_{\rm ej}$. Therefore, we only include them in the analysis of BH masses corresponding to $f_{\rm ej} \geqslant 0.5$. Otherwise, they would form BHs more than 1 $M_\odot$ smaller compared to the other stars, which would significantly impact the fit, despite these stars having an extremely low contribution to the full population due to the very small IMF at such high ZAMS masses. For $f_{\rm ej} \geqslant 0.5$ many of the other stars will produce BHs with masses comparable to or lower than $6 M_\odot$, and we can include these two progenitors in the analysis since they do not skew the fit at low masses anymore.

After calculating the BH masses according to procedures (i) and (ii) outlined above, we generate 20,000 synthetic black holes by weighting each BH mass with a Salpeter IMF with an exponent $\alpha=-2.35$. Then, using the "fit" method from the python module \texttt{statsmodels} \citep{statsmodels2010_py_package} we compute the best-fit distribution by maximizing the log-likelihood. The best-fit truncated PL and Gaussian distributions are shown in Figure \ref{fig:BH_distr} for different values of $f_{\rm ej}$ and for each of the metallicities, as well as for all of the metallicities combined. For simplicity, we only analyze the distribution of BHs for all of the metallicities combined due to the small samples of BHs produced at each given metallicity: 45, 32, and 7 for the solar, low, and zero metallicity sets, respectively. Therefore, for the remainder of this section, we always refer to the full set of the combined metallicities, unless stated otherwise. 

The best-fit parameters are summarized in Table \ref{tab:BH_best_fit} for both the BH masses calculated from simulations (i.e., procedure (i) described above, labeled as \texttt{GR1D+}) and the ones derived from our explodability criterion (i.e., procedure (ii) described above, labeled as pre-SN criterion). Table \ref{tab:ML_KS_BH} shows the value of the maximum log-likelihood, as well as the value of the KS test and the corresponding p-value for all of the fits. The p-value for the truncated PL is always $> 0.05$, whereas the Gaussian distribution has a p-value consistently lower than 0.01. Therefore, we conclude that a truncated PL is best to describe the data. This is not at all surprising since the pre-supernova mass grows approximately linearly with ZAMS mass, and therefore the low-end of the BH mass distribution will be truncated by construction, and then the power-law behavior is a consequence of the power-law distributed weights from the IMF.

A value of $f_{\rm ej} \gtrsim 0.8$ provides the closest distribution to what has been observed by LIGO. However, as mentioned above, the uncertainty in the masses of GW230529 is quite large, due to the low signal-to-noise ratio caused by the low amplitude of the gravitational waves from this event and the fact that only LIGO Livingston was online at the time. 

Previous studies on the remnant mass distributions have derived different values for $f_{\rm ej}$. For example, \cite{Ebinger2020_PUSH} adopt $f_{\rm ej} = 1$, and in some cases argue for the He-shell to be ejected as well, whereas \cite{Raithel2018_remnant_mass} find $f_{\rm ej} = 0.9$. Other studies \citep{Fryer2012_remnant_popsynth,Mandel2020_recipes_remn_kick} use different prescriptions where $f_{\rm ej}$ might be different for different progenitors and, in some cases, a fraction of the He-shell could also be ejected. 

\begin{figure}
\includegraphics[width=\columnwidth]{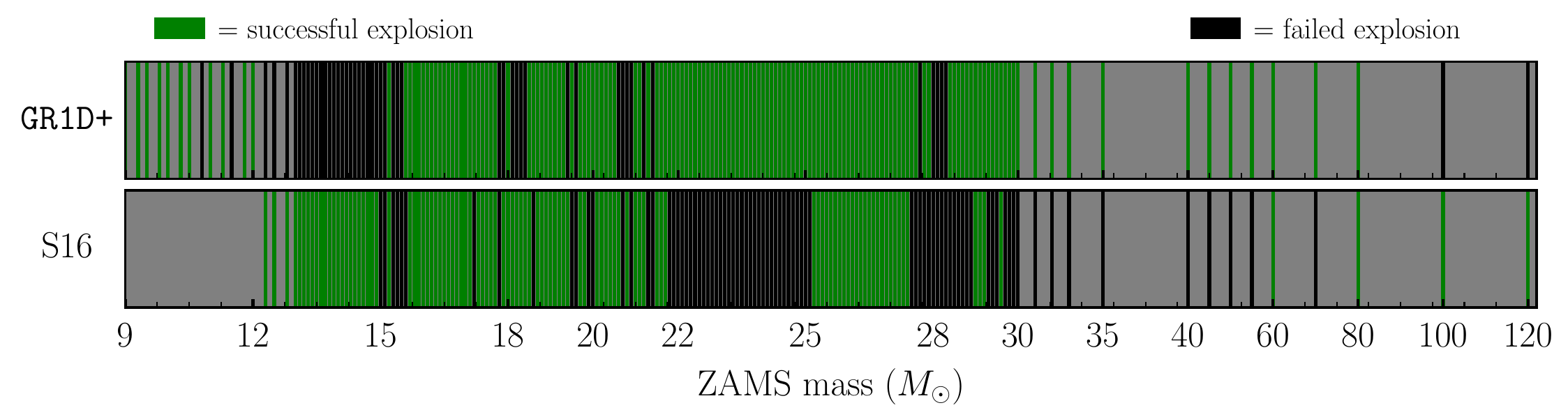}
\caption{\label{fig:N20_vs_STIR}Comparison between the explodability found using \texttt{GR1D+} simulations (upper panel) versus the explodability found by \cite{Sukhbold2016_explodability}. The initial pre-SN progenitors are, in both cases, the ones used in \citep{Sukhbold2016_explodability}. Notice that the upper panel is the same upper panel of the Solar Metallicity set shown in Figure \ref{fig:expl_pattern}.}
\end{figure}

Broadly speaking, our simulations produce lower-mass BHs compared to previous studies. This is a direct consequence of the completely different explodability pattern compared to previous studies. As an example, Figure \ref{fig:N20_vs_STIR} shows the difference between the outcomes of the simulations carried out in this work versus the ones from \cite{Sukhbold2016_explodability}. The pre-SN progenitors simulated in the two studies are exactly the same, but in our case, the explosion is achieved using the STIR, a physically motivated 1D+ model calibrated on multi-dimensional simulations, described in section \ref{sec:STIR_description}. Instead, \cite{Sukhbold2016_explodability} used an \textit{ad-hoc} prescription to manually change the neutrino luminosity emitted from the PNS calibrated on observed properties of supernovae \citep{Ugliano2012}. The other studies mentioned above \citep{Muller2016_prog_connection,Fryer2012_remnant_popsynth,Ebinger2020_PUSH} tend to have explodabilities that overall resemble \cite{Sukhbold2016_explodability}. On the other hand, 1D+ simulations from \cite{Couch2020_STIR} and 2D simulations from \cite{Wang2022_prog_study_ram_pressure} resemble the explodability pattern obtained by our simulations. For a more detailed discussion of these differences, we refer the reader to the discussion given in BR23.

As can be seen by the strikingly different explodability pattern in Figure \ref{fig:N20_vs_STIR}, our simulations produce black holes for ZAMS masses $12\ M_\odot < M_{\rm ZAMS} < 15\ M_\odot$, and do not produce black holes for ZAMS masses $22\ M_\odot < M_{\rm ZAMS} < 25\ M_\odot$, whereas \cite{Sukhbold2016_explodability} predict the exact opposite. Therefore, the black holes obtained in this work are much smaller. Hydrodynamic simulations \citep{Lovegrove2013_failed_SN_fej,Lovegrove2017_failedSN_fej_sims,Fernandez2018_failedSN_fej_progs,Ivanov2021_failed_SN_fej_EOS_nu,Schneider_2023_failed_SN_fej,Antoni2023_failed_SNeII} seem to consistently indicate that the hydrogen envelope is most likely ejected in the vast majority of cases, which would suggest $f_{\rm ej} \sim 1.0$. According to our simulations, this would correspond to a minimum BH mass of $\sim 2.6$ $M_\odot$, which corresponds to the mass of the smaller object in the merger event GW190814 \citep{Abbott2020_2.6_object}. Therefore, this would imply a continuous transition between NSs and BHs, i.e. no "low-mass gap" between NSs and BHs, which is what the most recent LVK results \citep{LIGO2024_GW230529} show. 

As was also mentioned in section \ref{sec:remnant_NS}, different stellar evolution codes predict different pre-supernova structures and, as a consequence, different BH and NS mass distributions. In particular, as shown in BR23, for the stellar evolution code FRANEC \citep{Chieffi2020_presupernova_models} our simulations show that black holes are formed at around $13-14$ $M_\odot$ and in the region between 20 $M_\odot$ and 24 $M_\odot$, which would modify the resulting BH and NS mass distributions. An investigation of how different stellar evolution codes and prescriptions affect the mass distributions of compact remnants is beyond the scope of this work and is left for future work.

Finally, it should be stressed that some of the most common prescriptions used for population synthesis calculations, such as \cite{Fryer2012_remnant_popsynth} and \cite{Patton2020_popsynth_prescription}, do not predict any black holes in the low-mass gap. Our model, instead, naturally explains the presence of low-mass black holes for values of $f_{\rm ej}$ close to 1, which is what the most recent, although still uncertain, theory seems to indicate. It is worth pointing out that another possible channel for creating low-mass black holes is if a high-compactness progenitor would initially launch an explosion and then later form a black hole due to late-time fallback. This has been recently seen by \cite{Burrows2023_BH_supernova_40Msol}, where an explosion of a 40 $M_\odot$ progenitor yielded a successful explosion followed by the formation of a $\sim 2.63\ M_\odot$ black hole. This is still a relatively unexplored phenomenon, due to its complexity and computational demand, and it is therefore unclear if black holes with masses $\sim 3-5\ M_\odot$ could be produced that way. For now, as we have shown, the black hole formation channel presented in this work can naturally explain a plethora of black holes at low masses.

To summarize, the remnant masses predicted by our simulations are in excellent agreement with the most recent LVK results \citep{LIGO2024_GW230529}, which seem to indicate that there is no mass gap, and therefore there is a smooth transition between NSs and BHs. To obtain such low-mass stellar black holes, values of $f_{\rm ej} > 0.8$ are required, which is what the current models of failed SNe suggest. Given the large uncertainty in the masses of GW230529, the exact location of the low-mass cutoff for stellar black holes is still not well constrained, and could be anywhere within the range $2.5-4.5\ M_\odot$, and therefore all values of $f_{\rm ej} > 0.8$ are plausible. The difference between the BH masses derived using the explodability found directly from the simulation and the BH masses derived assuming the explodability predicted by the criterion described in section \ref{sec:criterion} is very small. Therefore, one can robustly predict the outcome of the supernova and the mass of the resulting BH (after choosing the value of $f_{\rm ej}$) simply based on the pre-SN density and entropy profiles.

\begin{table*}
\centering 
\begin{ruledtabular}
\begin{tabular}{>{\centering\arraybackslash}p{25mm}|ccccccccccc}

\multicolumn{12}{c}{Gaussian Fit} \\
\midrule
\midrule
\multicolumn{12}{c}{$\mu$} \\
\midrule
{$f_{\rm ej}$} &    0.0 &    0.1 &    0.2 &    0.3 &   0.4 &   0.5 &   0.6 &   0.7 &   0.8 &   0.9 &   1.0 \\
\midrule
\texttt{GR1D+}   &  13.50 &  12.55 &  11.60 &  10.69 &  9.78 &  8.85 &  7.91 &  7.00 &  6.08 &  5.16 &  4.23 \\
pre-SN criterion &  13.49 &  12.55 &  11.60 &  10.70 &  9.75 &  8.79 &  7.90 &  6.94 &  6.02 &  5.09 &  4.16 \\
\midrule
\multicolumn{12}{c}{$\sigma$} \\
\midrule
{$f_{\rm ej}$} &    0.0 &    0.1 &    0.2 &    0.3 &   0.4 &   0.5 &   0.6 &   0.7 &   0.8 &   0.9 &   1.0 \\
\midrule
\texttt{GR1D+}   &   2.92 &   2.65 &   2.49 &   2.34 &  2.20 &  2.03 &  1.86 &  1.72 &  1.61 &  1.46 &  1.35 \\
pre-SN criterion &   2.95 &   2.79 &   2.59 &   2.40 &  2.23 &  2.00 &  1.91 &  1.69 &  1.55 &  1.41 &  1.28 \\
\midrule
\midrule
\multicolumn{12}{c}{Truncated PL Fit} \\
\midrule
\midrule
\multicolumn{12}{c}{m$_{\rm min}$} \\
\midrule
{$f_{\rm ej}$} &    0.0 &    0.1 &    0.2 &    0.3 &   0.4 &   0.5 &   0.6 &   0.7 &   0.8 &   0.9 &   1.0 \\
\midrule
\texttt{GR1D+}   &   9.58 &   8.80 &   8.04 &   7.32 &  6.63 &  5.81 &  5.22 &  4.44 &  3.65 &  2.87 &  2.09 \\
pre-SN criterion &   9.90 &   9.12 &   8.34 &   7.56 &  6.78 &  5.98 &  5.22 &  4.44 &  3.65 &  2.87 &  2.09 \\
\midrule
\multicolumn{12}{c}{$\alpha$} \\
\midrule
{$f_{\rm ej}$} &    0.0 &    0.1 &    0.2 &    0.3 &   0.4 &   0.5 &   0.6 &   0.7 &   0.8 &   0.9 &   1.0 \\
\midrule
\texttt{GR1D+}   &   6.11 &   6.09 &   6.01 &   6.08 &  6.21 &  5.88 &  6.57 &  6.29 &  6.10 &  5.76 &  5.33 \\
pre-SN criterion &   7.52 &   7.42 &   7.63 &   7.20 &  7.01 &  6.68 &  6.74 &  6.47 &  6.21 &  5.80 &  5.42 \\

\end{tabular}
\end{ruledtabular}

\caption{Best-fit parameters for the Gaussian and truncated power-law fits for the two sets of theoretical BH masses described in section \ref{sec:remnant_BH}, for different values of $f_{\rm ej}$. \label{tab:BH_best_fit}}

\end{table*}

\begin{table*}
\centering 

\begin{ruledtabular}
\begin{tabular}{>{\centering\arraybackslash}p{25mm}|ccccccccccc}

\multicolumn{12}{c}{Gaussian Fit} \\
\midrule
\midrule
\multicolumn{12}{c}{MLL} \\
\midrule
{$f_{\rm ej}$} &    0.0 &    0.1 &    0.2 &    0.3 &   0.4 &   0.5 &   0.6 &   0.7 &   0.8 &   0.9 &   1.0 \\
\midrule
\texttt{GR1D+}   &  -49779 &  -47885 &  -46634 &  -45407 &  -44104 &  -42566 &  -40779 &  -39265 &  -37904 &  -35943 &   -34347 \\
pre-SN criterion &  -49982 &  -48866 &  -47383 &  -45923 &  -44388 &  -42269 &  -41343 &  -38921 &  -37127 &  -35186 &   -33267 \\
\midrule
\multicolumn{12}{c}{KS test} \\
\midrule
{$f_{\rm ej}$} &    0.0 &    0.1 &    0.2 &    0.3 &   0.4 &   0.5 &   0.6 &   0.7 &   0.8 &   0.9 &   1.0 \\
\midrule
\texttt{GR1D+}   &    0.22 &    0.22 &    0.23 &    0.23 &    0.23 &    0.23 &    0.23 &    0.23 &    0.23 &    0.25 &     0.30 \\
pre-SN criterion &    0.26 &    0.26 &    0.26 &    0.26 &    0.26 &    0.26 &    0.27 &    0.27 &    0.27 &    0.28 &     0.33 \\
\midrule
\multicolumn{12}{c}{p-value} \\
\midrule
{$f_{\rm ej}$} &    0.0 &    0.1 &    0.2 &    0.3 &   0.4 &   0.5 &   0.6 &   0.7 &   0.8 &   0.9 &   1.0 \\
\midrule
\texttt{GR1D+}   &   0.024 &   0.020 &   0.018 &   0.017 &   0.016 &   0.016 &   0.014 &   0.015 &   0.016 &  0.0057 &  0.00056 \\
pre-SN criterion &   0.015 &   0.015 &   0.017 &   0.017 &   0.016 &   0.012 &   0.011 &  0.0092 &   0.010 &  0.0072 &  0.00059 \\
\midrule
\midrule
\multicolumn{12}{c}{Truncated PL Fit} \\
\midrule
\midrule
\multicolumn{12}{c}MLL \\
\midrule
{$f_{\rm ej}$} &    0.0 &    0.1 &    0.2 &    0.3 &   0.4 &   0.5 &   0.6 &   0.7 &   0.8 &   0.9 &   1.0 \\
\midrule
\texttt{GR1D+}   &  -40715 &  -39580 &  -38495 &  -37280 &  -36133 &  -35107 &  -34080 &  -32807 &  -31657 &  -30653 &   -30545 \\
pre-SN criterion &  -38311 &  -37400 &  -36385 &  -35367 &  -34307 &  -33266 &  -31796 &  -30755 &  -29854 &  -29301 &   -29055 \\
\midrule
\multicolumn{12}{c}{KS test} \\
\midrule
{$f_{\rm ej}$} &    0.0 &    0.1 &    0.2 &    0.3 &   0.4 &   0.5 &   0.6 &   0.7 &   0.8 &   0.9 &   1.0 \\
\midrule
\texttt{GR1D+}   &    0.15 &    0.16 &    0.15 &    0.15 &    0.15 &    0.16 &    0.15 &    0.15 &    0.15 &    0.16 &     0.16 \\
pre-SN criterion &    0.19 &    0.18 &    0.19 &    0.18 &    0.18 &    0.19 &    0.19 &    0.20 &    0.20 &    0.20 &     0.21 \\
\midrule
\multicolumn{12}{c}{p-value} \\
\midrule
{$f_{\rm ej}$} &    0.0 &    0.1 &    0.2 &    0.3 &   0.4 &   0.5 &   0.6 &   0.7 &   0.8 &   0.9 &   1.0 \\
\midrule
\texttt{GR1D+}   &    0.25 &    0.21 &    0.22 &    0.22 &    0.24 &    0.19 &    0.21 &    0.23 &    0.22 &    0.19 &     0.16 \\
pre-SN criterion &    0.16 &    0.18 &    0.17 &    0.18 &    0.18 &    0.13 &    0.13 &    0.12 &    0.12 &    0.11 &    0.084 \\

\end{tabular}
\end{ruledtabular}

\caption{Properties of the best-fit distributions for the two sets of theoretical BH masses described in section \ref{sec:remnant_BH}, for different values of $f_{\rm ej}$. The MLL is the Maximum Log-Likelihood of the best-fit distribution. The KS-test is the value of the Kolmogorov-Smirnov test, and the p-value is the probability of rejecting the null hypothesis that the raw data are not drawn from the fitted distribution. To perform the KS test, we binned the BH masses by choosing a number of equal-width bins equal to half of the total number of BH masses in each of the two sets of theoretical BH masses. A more detailed description of the procedure is given in appendix \ref{sec:appendix_BH_fit}. \label{tab:ML_KS_BH}}
\end{table*}

\section{Summary and Conclusions}
\label{sec:conclusions}
In this paper, we have analyzed the remnant masses of 341 1D+ simulations for progenitors spanning a wide range of ZAMS masses and metallicities. The explodability of CCSNe remains a challenging problem, affected by uncertainties in the supernova engine itself as well as by uncertainties in the stellar evolution \citep{Boccioli2024_Review}. We showed that 1D+ simulations \citep{Couch2020_STIR,Boccioli2023_explodability} predict a very different explodability pattern compared to previous 1D studies \citep{Ertl2016_explodability,Muller2016_prog_connection,Ebinger2020_PUSH,Fryer2012_remnant_popsynth}. However, they are consistent with the most recent 2D and 3D simulations \citep{Burrows2020_3DFornax,Wang2022_prog_study_ram_pressure}, due to the physically consistent (although parametric) model for $\nu$-driven convection implemented.

We provide a robust criterion to predict the explodability of CCSNe based on the pre-SN density and entropy profiles. This criterion is a modification of the one derived by BR23, to account for the exploding high-compactness progenitors at low and zero metallicity which, according to the simpler criterion of BR23, are predicted to produce a failed SN.

The masses of neutron stars produced by the explosion of CCSNe are highly dependent on the pre-SN structure of the massive star in question. In particular, there is a positive correlation between the Chandrasekhar mass and the gravitational mass of the final, cold neutron star. Generally, the Chandrasekhar mass is always slightly smaller than the mass of the neutron star. The dependence of the Chandrasekhar mass $M_{\rm Ch}$ on the ZAMS mass of the star is nontrivial, and affected by several uncertainties. Broadly speaking, one expects stars above a certain ZAMS mass to burn Carbon in radiative equilibrium, rather than in convective equilibrium. This eventually causes the star to completely bypass the neutrino cooling phase in the latest stages its life. This prevents the central entropy from decreasing, and therefore one naturally expects a bimodal distribution of $M_{\rm Ch}$. This would therefore translate into a bimodal distribution of NS masses \citep{Timmes1996_NS_BH_birth_mass}. However, most previous 1D studies found that the stars that burn Carbon in radiative equilibrium do not successfully explode, but instead lead to the formation of a black hole. Therefore, most of those studies found a single-peaked NS mass distribution. Our 1D+ simulations, however, find that these stars (e.g. stars with ZAMS masses $22\ M_\odot \lesssim M \lesssim 25\ M_\odot$ at solar metallicity, according to the KEPLER stellar evolution code \citep{Sukhbold2016_explodability}) successfully explode, in agreement with 2D and 3D simulations. 

In light of this novel result, we analyzed the remnant masses from our 1D+ simulations and derived four theoretical distributions. The first is calculated directly from the \texttt{GR1D+} simulations. The remaining three are calculated using the explodability criterion described in section \ref{sec:criterion}, and three different fits to compute the neutron star gravitational mass based on compactness, $M_{\rm Si/O}$, and $M_{\rm Ch}$. In all cases, we found a robust second peak at $\sim 1.8\ M_\odot$, compatible with the observed population of ms pulsars from \cite{Antoniadis2016_NS_mass_distr}. This could potentially have a significant impact on binary population synthesis calculations, since one would be able to produce very massive neutron stars directly from a SN, without the need to accrete large amounts of matter via binary interactions.

When compared to previous 1D simulations, our \texttt{GR1D+} simulations also produce lower mass black holes (e.g. for stars with ZAMS masses $12\ M_\odot \lesssim M \lesssim 15\ M_\odot$ at solar metallicity, according to the KEPLER stellar evolution code \citep{Sukhbold2016_explodability}). We derived black hole mass distributions directly from our simulations and also using the explodability criterion described in section \ref{sec:criterion}. Given the very high accuracy of our explodability criterion, the two theoretical distributions are in excellent agreement. In all cases, a truncated power-law fits the data much better than a Gaussian. This is not surprising and is simply due to the linear dependence of the BH mass on the ZAMS mass, and the power-law behavior of the IMF.

We then compared these results with the BH mass distribution observed by the LVK collaboration, in light of the very recent event GW230529 \citep{LIGO2024_GW230529}. A black hole in the so-called "low-mass gap" was observed, which is naturally explained by our simulations if one assumes that more than 80 \% (and up to 100 \%) of the hydrogen envelope is expelled. Indeed, there is a general consensus \citep{Lovegrove2013_failed_SN_fej,Lovegrove2017_failedSN_fej_sims,Fernandez2018_failedSN_fej_progs,Ivanov2021_failed_SN_fej_EOS_nu,Schneider_2023_failed_SN_fej,Antoni2023_failed_SNeII} that most (if not the entirety) of the hydrogen envelope will be ejected, which corresponds to the range of $f_{\rm ej}$ that we found. This can have a significant impact on binary population synthesis calculations, since many more black holes (although with a smaller mass) compared to predictions of previous 1D studies will be produced, since stars with ZAMS masses $12\ M_\odot \lesssim M \lesssim 15\ M_\odot$ are more favored by the IMF. However, it is important to stress that the fraction of failed SNe is still compatible with the most recent (and more stringent) observational data, even though these 1D+ simulations produce black holes for stars that are favored by the IMF.

In conclusion, the explosion properties of CCSNe are still affected by several uncertainties, and the explodability of CCSNe is still a highly debated topic. More sophisticated 3D (and, although less reliable, 2D) simulations are required to shed more light on this complicated phenomenon. With that in mind, we have presented 1D+ simulations that are in very good agreement with the most recent sets of 2D and 3D simulations. We used these simulations to derive a simple explodability criterion that generalizes the one derived by BR23. Moreover, we derived simple fits to calculate the remnant masses of neutron stars and black holes solely based on the pre-SN structure of massive stars, regardless of metallicity. The two main findings of this paper are the bimodal nature of NS birth mass distribution, as well as the presence of many black holes in the low-mass gap. Both of these hypotheses are completely viable, and future observations will be able to put better constraints on both of them, with the latter being already confirmed by the most recent results \citep{LIGO2024_GW230529} of the LVK collaboration.

\section*{Acknowledgements}
L.B. would like to thank David Vartanyan and Feryal \"Ozel for fruitful discussions, and the N3AS center for their hospitality and support. L.B. was supported by the U.S. Department of Energy under Grant No. DE-SC0004658.


\bibliography{References_Books,References_CNO,References_EoS_neutrinos,references_misc,References_Exp_Obs,References_Nucleosynthesis,References_SN,References_Stellar_Models}



\appendix
\section{Fitting procedure for NS masses}
\label{sec:appendix_NS_distr}
The fitting procedure to derive NS mass distributions is as follows: 
\begin{itemize}
    \item select the successful explosions directly from the simulations (set (i) in section \ref{sec:remnant_NS_distr}) or by using the explodability criterion described in section \ref{sec:criterion} (sets (ii), (iii), and (iv) in section \ref{sec:remnant_NS_distr});
    \item calculate the NS mass by converting the final baryonic mass of the PNS to gravitational mass (set (i) in section \ref{sec:remnant_NS_distr}) or by using the $\xi$-fit, $M_{\rm Si/O}$-fit, or $M_{\rm Ch}$-fit (sets (ii), (iii), and (iv), in section \ref{sec:remnant_NS_distr}) described in equations \eqref{eq:comp_fit} and \eqref{eq:broken_pl};
    \item generate $20000$ synthetic observations by randomly sampling the NS masses weighted by an IMF $\propto M_{\rm ZAMS}^{-2.35}$;
    \item perform a ML fit using the python module \texttt{statsmodels} \citep{statsmodels2010_py_package}.
\end{itemize}
Notice that, instead of using the final baryonic mass of the PNS as a proxy for the final mass of the cold NS, one could also define a mass cut $M_{\rm cut}$, as done for example in \cite{Perego2015_PUSH1}. Below $M_{\rm cut}$, matter will accrete onto the compact object, and above $M_{\rm cut}$ matter will instead be ejected. However, the mass cut is located extremely close (in mass) to the PNS, and we did not find any appreciable differences in the remnant mass defined in these two different ways.

\begin{figure*}
\includegraphics[width=\textwidth]{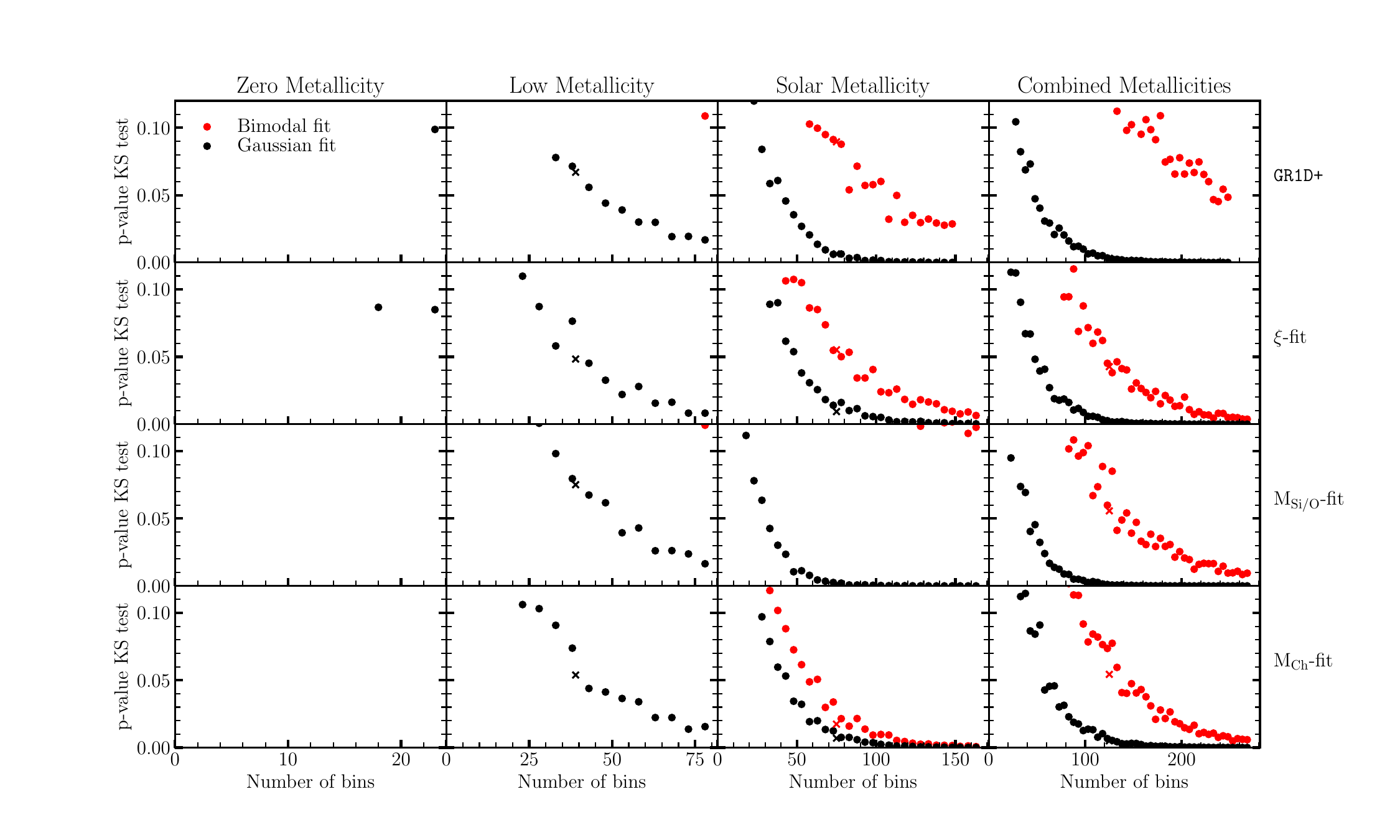}
\caption{Probability of rejecting the null hypothesis (p-value) for the Kolmogorov–Smirnov test as a function of the number of equal-width bins used to bin the data. The KS test has been performed between the four fitted distributions described in section \ref{sec:remnant_NS_distr} (i.e. the four rows in the plot) and the raw data, for different metallicities and all metallicities combined (i.e. the four columns in the plot). Values below 0.05 indicate that the data is not distributed according to the fitted distribution. Red-filled circles represent the p-value calculated assuming a bimodal theoretical distribution; black-filled circles represent the p-value calculated assuming a Gaussian theoretical distribution. Red (black) crosses indicate the values reported in Table \ref{tab:ML_KS_NS} for the bimodal (Gaussian) fit. \label{fig:NS_pval_vs_nbins}}
\end{figure*}

To perform the KS test to assess the goodness-of-fit, one has to first bin the data. The results reported in table \ref{tab:ML_KS_NS} were derived assuming a number of bins equal to half of the total number of successful simulations. For completeness, we show in Figure \ref{fig:NS_pval_vs_nbins} how the p-value for the KS test changes as a function of the number of equal-width bins chosen. As one would intuitively expect, with only a few bins the KS test cannot reject the null hypothesis.

\begin{figure*}
\includegraphics[width=\textwidth]{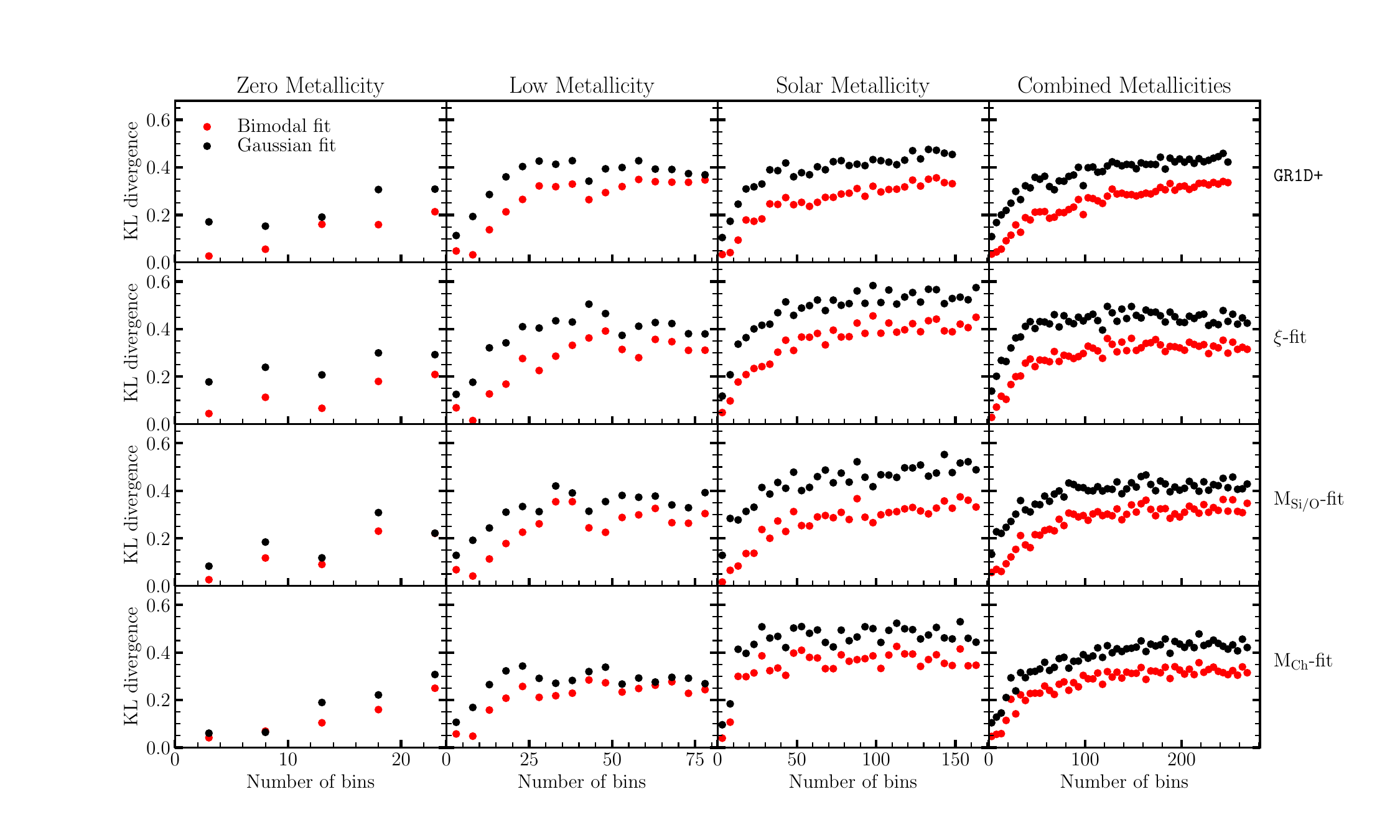}
\caption{Kullback–Leibler divergence computed between the four fitted distributions described in section \ref{sec:remnant_NS_distr} (i.e. the four rows in the plot) and the raw data as a function of the number of equal-width bins used to bin the data, for different metallicities and all metallicities combined (i.e. the four columns in the plot). Red-filled circles represent the KL divergence calculated between a bimodal theoretical distribution and the raw data; black-filled circles represent the KL divergence calculated between a Gaussian theoretical distribution and the raw data \label{fig:NS_KL_vs_nbins}}
\end{figure*}

Due to the potential ambiguity of the KS test in this scenario, we also computed the Kullback–Leibler (KL) divergence between the fitted distribution and the actual distribution of the data. This metric essentially determines how "close" two distributions are. Similarly to the KS test, we first need to bin the data to perform the comparison, and we show in Figure \ref{fig:NS_KL_vs_nbins} that, regardless of this choice, the bimodal distribution is always a closer representation of the data.

\section{Fitting procedure for BH masses}
\label{sec:appendix_BH_fit}
The fitting procedure to derive BH mass distributions is as follows: 
\begin{itemize}
    \item select the successful explosions directly from the simulations (set (i) in section \ref{sec:remnant_BH}) or by using the explodability criterion described in section \ref{sec:criterion} (set (ii) in section \ref{sec:remnant_NS_distr});
    \item calculate the BH mass by ejecting a fraction $f_{\rm ej}$ of the hydrogen envelope (eq. \eqref{eq:MBH};
    \item generate $20000$ synthetic observations by randomly sampling the BH masses weighted by an IMF $\propto M_{\rm ZAMS}^{-2.35}$;
    \item perform a ML fit using the python module \texttt{statsmodels} \citep{statsmodels2010_py_package}.
\end{itemize}
In this approach, $f_{\rm ej}$ is essentially almost a free parameter, although some theoretical studies suggest values closer to 1 as more plausible \citep{Lovegrove2013_failed_SN_fej,Lovegrove2017_failedSN_fej_sims,Fernandez2018_failedSN_fej_progs,Ivanov2021_failed_SN_fej_EOS_nu,Schneider_2023_failed_SN_fej,Antoni2023_failed_SNeII}.

\begin{figure*}
\includegraphics[width=\textwidth]{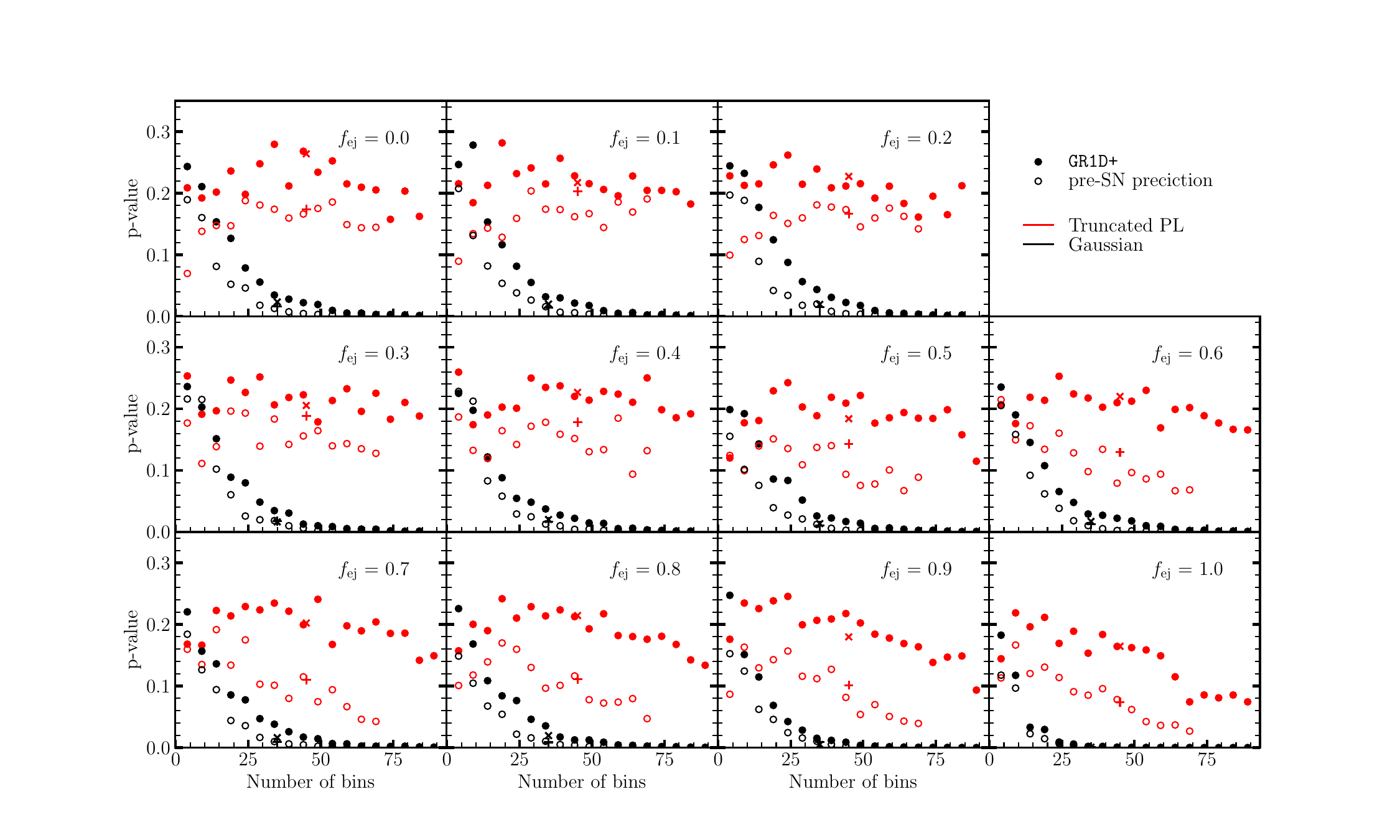}
\caption{Probability of rejecting the null hypothesis (p-value) for the Kolmogorov–Smirnov test as a function of the number of equal-width bins used to bin the data. The KS test has been performed between the two fitted distributions described in section \ref{sec:remnant_BH} (shown as filled or empty circles) and the raw data for different values of $f_{\rm ej}$. Values below 0.05 indicate that the data is not distributed according to the fitted distribution. Red circles represent the p-value calculated assuming a bimodal theoretical distribution; black circles represent the p-value calculated assuming a Gaussian theoretical distribution. Crosses and plus symbols indicate the values reported in Table \ref{tab:ML_KS_BH} for fit performed using the explodability obtained in the simulations and the explodability predicted by the pre-SN explodability criterion described in section \ref{sec:criterion}, respectively. \label{fig:BH_pval_vs_nbins}}
\end{figure*}

\begin{figure*}
\includegraphics[width=\textwidth]{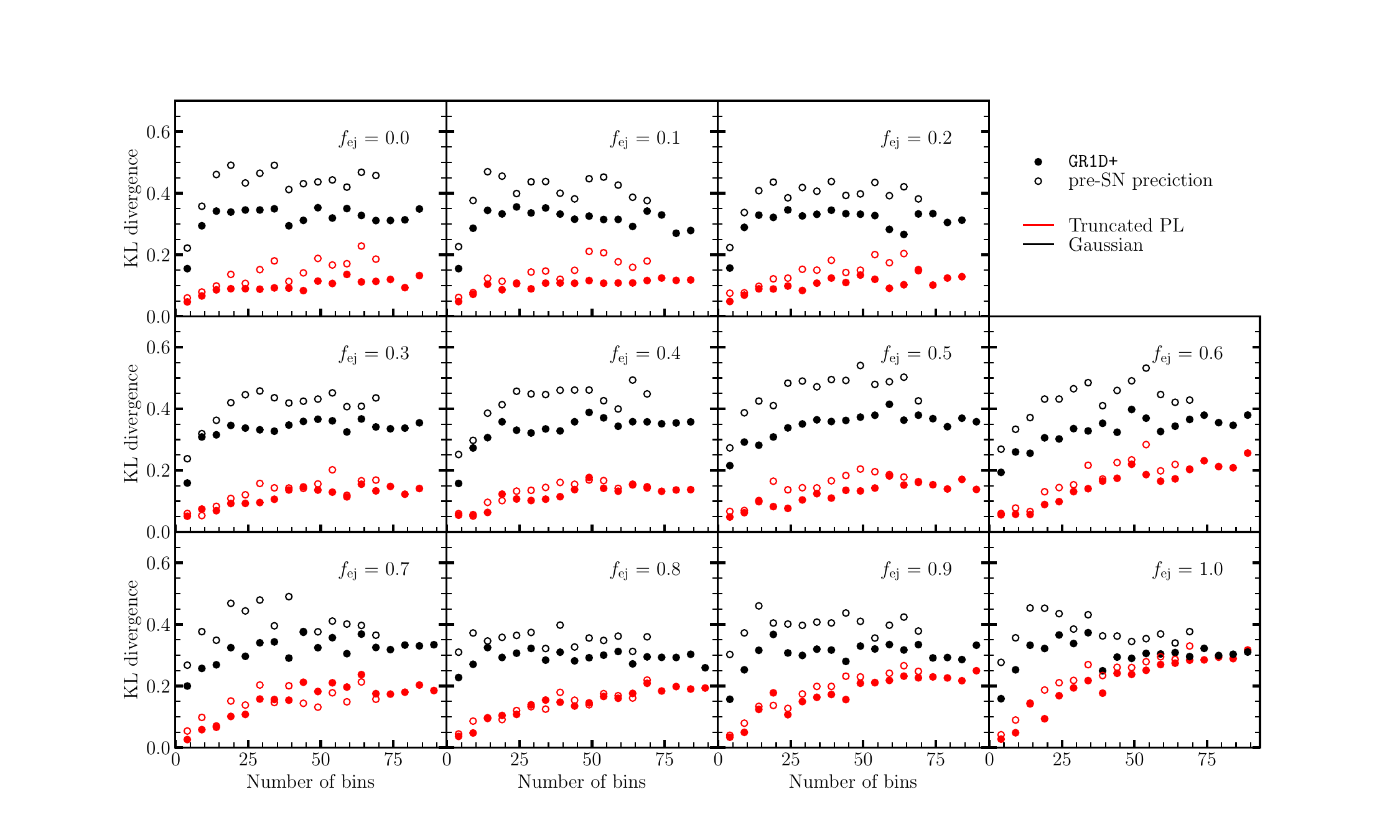}
\caption{Kullback–Leibler divergence computed between the two fitted distributions described in section \ref{sec:remnant_NS_distr} (shown as filled or empty circles) and the raw data as a function of the number of equal-width bins used to bin the data, for different values of $f_{\rm ej}$. Red circles represent the KL divergence calculated between a bimodal theoretical distribution and the raw data; black circles represent the KL divergence calculated between a Gaussian theoretical distribution and the raw data. \label{fig:BH_KL_vs_nbins}}
\end{figure*}

Due to the small samples of data compared to the case of NSs, we only show the results of the KS test and KL divergence for the combined metallicities case, for different values of $f_{\rm ej}$ in Figures \ref{fig:BH_pval_vs_nbins} and \ref{fig:BH_KL_vs_nbins}. The procedure is the same followed in the case of NSs.

\end{document}